\documentclass[pre, twocolumn, showpacs, preprintnumbers,amsmath,amssymb,superscriptaddress]{revtex4-1}\usepackage[utf8]{inputenc}
\usepackage{bm}
\usepackage{verbatim}
\usepackage{amsmath}
\usepackage{amssymb}
\usepackage{amsfonts}
\usepackage{mathtools}
\usepackage{dcolumn}
\usepackage{graphicx}
\usepackage{cancel}
\usepackage{color,soul}
\usepackage[english]{babel}
\usepackage{csquotes}
\usepackage[title]{appendix}
\usepackage{array}
\usepackage{multirow}
\usepackage{makecell}
\usepackage[title]{appendix}
\newcommand{\dd}{\mathrm{d}}

\begin{document}

\title{Dynamic assembly of active colloids: theory and simulation}

\author{Zhan Ma}
\affiliation{School of Chemical and Biomedical Engineering, Nanyang Technological University, 62 Nanyang Drive, 637459, Singapore}

\author{Mingcheng Yang}
\email{mcyang@iphy.ac.cn}
\affiliation{Beijing National Laboratory for Condensed Matter Physics and Institute of Physics, Chinese Academy of Sciences, Beijing 100190, China}
\affiliation{School of Physical Sciences, University of Chinese Academy of Sciences, Beijing 100049, China}

\author{Ran Ni}
\email{r.ni@ntu.edu.sg}
\affiliation{School of Chemical and Biomedical Engineering, Nanyang Technological University, 62 Nanyang Drive, 637459, Singapore}

\begin{abstract}
Because of consuming energy to drive their motion, systems of active colloids are intrinsically out of equilibrium. In the past decade, a variety of intriguing dynamic patterns have been observed in systems of active colloids, and they offer a new platform for studying non-equilibrium physics, in which computer simulation and analytical theory have played an important role. Here we review the recent progress in understanding the dynamic assembly of active colloids by using numerical and analytical tools. We review the progress in understanding the motility induced phase separation in the past decade, followed by the discussion on the effect of shape anisotropy and hydrodynamics on the dynamic assembly of active colloids.
\end{abstract}

\maketitle

\section{Introduction}
Active matter are essentially the particles or objects, that are capable of converting ambient or stored energy into their self-propulsion, and they are intrinsically out of equilibrium. The original purpose of investigating active matter was to understand the emergent behaviour in nature, like bird flocks, bacteria colonies, tissue repair, and cell cytoskeleton~\cite{activematterrev}. Recent breakthroughs in particle synthesis have produced a spectacular variety of novel building blocks, which offers new possibilities to fabricate synthetic artificial active colloids, whose dynamics can be better controlled for investigating the emergent behaviour of non-equilibrium active matter~\cite{partsyn}.
Various active colloidal systems have been realized in experiments, such as colloids with magnetic beads acting as artificial flagella~\cite{swimm1}, catalytic Janus particles~\cite{swimm2,swimm3,swimm4,swimm5}, laser-heated metal-capped particles~\cite{swimm6}, light-activated catalytic colloidal surfers~\cite{palacci2013living}, and platinum-loaded stomatocytes~\cite{swimm7}. In contrast to passive colloids undergoing Brownian motion due to random thermal fluctuations of the solvent, active colloids experience an additional force due to internal energy conversion.
Although the long time dynamics of self-propelled particles is still Brownian, with a mean square displacement proportional to time~\cite{swimm8}, the self-propulsion has produced a variety of strikingly new phenomena, which were never observed in corresponding systems of passive particles~\cite{swimm9}, e.g., bacteria ratchet motors~\cite{swimm10}, mesoscale turbulence~\cite{swimm11}, living crystals~\cite{palacci2013living}, motility induced phase separation~\cite{cates2015motility}, emergent long range effective interactions~\cite{niprl2015,swimm12}, hyperuniform fluids~\cite{leihu2019sa}, etc. All these offered a new testbed for investigating non-equilibrium physics, in which theory and simulation have played an important role. Here we review the recent progress in understanding the emergent dynamic assembly of active colloids by using analytical mean field theories and computer simulations. Particularly, in Sec.~\ref{MIPSSEC}, we first briefly review the recent progress in understanding the motility induced phase separation in systems of active repulsive spheres, which is arguably the ``simplest'' active colloidal systems. In Sec.~\ref{AACSEC}, we show various dynamic patterns found in systems of anisotropic active colloids. In Sec.~\ref{HESEC}, we discuss the effect of hydrodynamics on the dynamic assembly of active colloids. Lastly, concluding remarks are given in Sec.~\ref{conclude}.

\section{Motility induced phase separation for active repulsive spheres}\label{MIPSSEC}
One of the most intriguing phenomena in active matter is the existence of the motility induced (gas-liquid like) phase separation (MIPS) without any attraction between particles~\cite{cates2015motility}, which was proven neccessary to induce the gas-liquid transition in equilibrium~\cite{vanderwaals}. The essential physics driving MIPS is the ``self-trapping'' effect first proposed by Tailleur and Cates using the model of run-and-tumble self-propelled particles in one dimension~\cite{selftrappingcates}, which was observed in simulations~\cite{fily2012athermal,redner2013structure} and experiments~\cite{lowenmips2013} of two dimensional self-propelled colloidal systems. Recently, in three dimensional systems of active hard spheres, MIPS has also been confirmed~\cite{mips3d}. For a comprehensive review on MIPS, please refer to Ref.~\cite{cates2015motility}, and in the following, we briefly review the recent progress on the theoretical understanding of MIPS.

\subsection{Active Brownian particles}
\begin{figure}[htbp]
\centering
{\includegraphics[width=1.\linewidth]{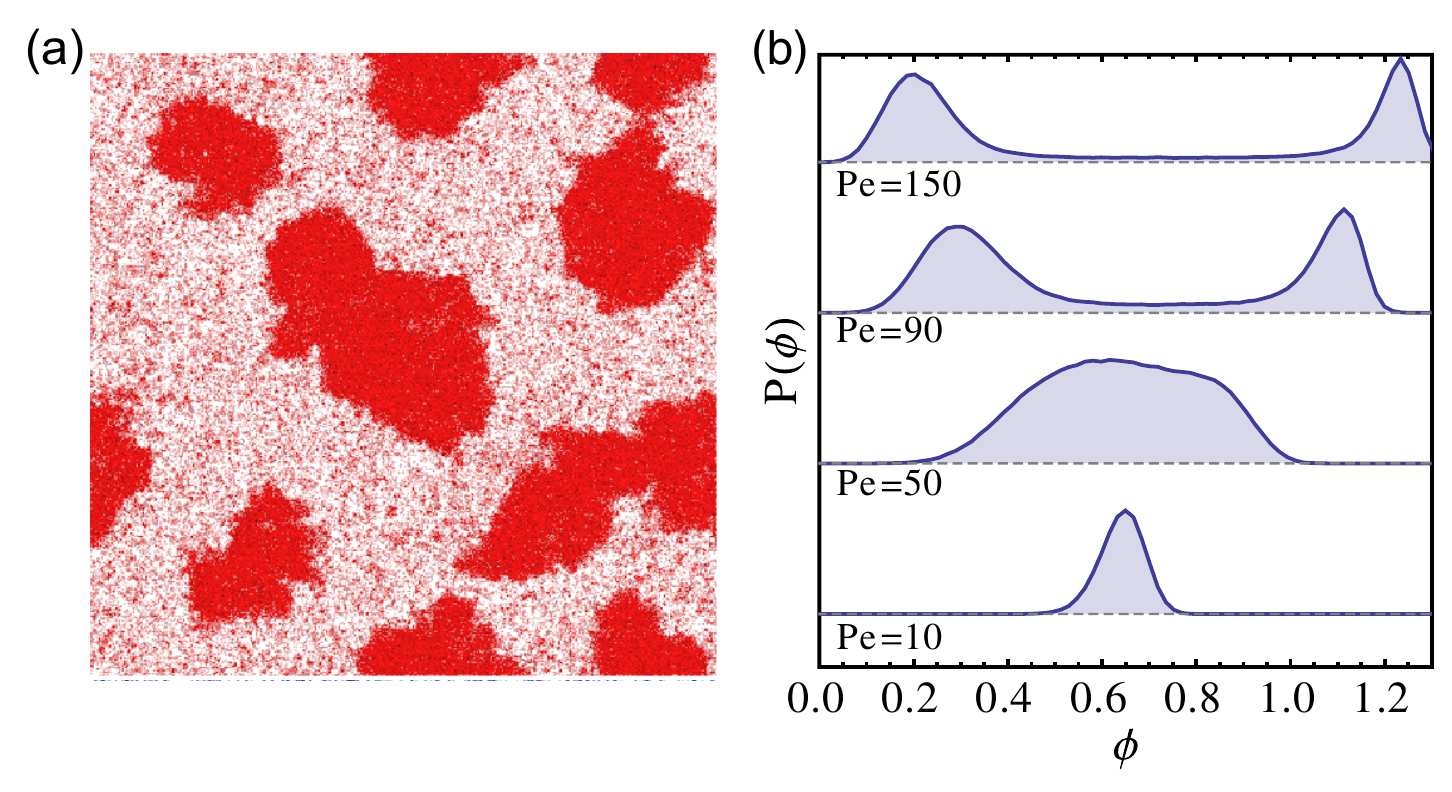}}
\caption{\label{fig1} (a) A typical snapshot for the system of ABPs beyond the critical density and activity levels that the ABP fluid separates into a dense and a dilute phase. (b) Observed density distributions for various P{\'{e}}clet numbers~\cite{redner2013structure}. 
Reproduced with permission.\cite{redner2013structure} Copyright 2013, American Physical Society.
}
\end{figure}

The most commonly used model for investigating MIPS is the system of active Brownian particles (ABPs), of which the physics can be also applied to some other self-propelled particle model systems, e.g., run-and-tumble particles~\cite{cates2015motility}.
Even though ABPs are driven and energy is continuously supplied to the system, the solvent is assumed to stay at a constant temperature $T$ acting as a heat bath, and the equation of motion for particle $i$ can be described via the overdamped Langevin equation:
\begin{eqnarray}
\dot{\mathbf{r}}_i(t) &=& \gamma_t^{-1} \left[-{\nabla}_i U(t) + F_p \textbf{e}_i(t)\right] + \sqrt{2 k_B T/\gamma_t}~{\bm \xi}^t_i(t), \label{eq1n}
\\
\dot{\textbf{e}}_i(t) &=& \sqrt{2 k_B T/\gamma_r}~{\bm \xi}^r_i(t)  \times {\textbf{e}}_i(t),\label{eq2}
\end{eqnarray}
where $\mathbf{r}_i$ and $\mathbf{e}_i$ are the position of particle $i$ and its self-propulsion orientation, respectively, and $F_p$ is the strength of self-propulsion with $k_B$ the Boltzmann constant. $\gamma_{t/r}$ is the translational/rotational friction coefficient.  ${\bm \xi}^t_i(t)$ and ${\bm \xi}^r_i(t)$ represent Gaussian white noises with zero mean and unit variance.
$U(t) = \sum_{i>j}U_{ij}$ is the potential of the system at time $t$, and the interaction between ABP $i$ and $j$ can be modelled as the hard-core interaction with diameter $\sigma$~\cite{ni2013natcomm,rni2014sm} or the Week-Chandler-Anderson (WCA) potential~\cite{redner2013structure}:
\begin{equation}
U(r_{ij}) = \left\{
\begin{array}{ll}
4\epsilon\left[\left(\frac{\sigma}{r_{ij}}\right)^{12} - \left(\frac{\sigma}{r_{ij}}\right)^6 + \frac{1}{4}\right] & (r_{ij} < 2^{1/6}\sigma),\\
0 & (r_{ij} \ge 2^{1/6}\sigma),
\end{array}\right.
\end{equation}
where $r_{ij}$ is the center-to-center distance between particle $i$ and $j$ with $\epsilon \simeq k_BT$.
It was recently confirmed that the soft repulsion in the WCA potential does not influence significantly the collective behaviour of ABPs~\cite{levis2017softmatt}.
In Ref.~\cite{redner2013structure}, a dimensionless P{\'{e}}clet number is defined as $\mathrm{Pe} = F_p \sigma /k_BT$ to characterize the effect of self-propulsion. It was found that in a 2D system of ABPs, with increasing Pe above certain threshold, the system phase separates into a dense (dynamic clusters) and a dilute phase (Fig.~\ref{fig1}a), and the probability distribution of local density becomes bimodal (Fig.~\ref{fig1}b). It was confirmed that the dynamic clusters merge into one single cluster in long enough simulations implying a first-order like phase separation~\cite{redner2013structure}. As the interaction between ABPs is purely repulsive, the phase separation is solely induced by the motility of ABPs, i.e., motility induced phase separation. It has been confirmed that the thermal noise ${\bm \xi}^t_i(t)$ on the translational degree of freedom does not qualitatively influence MIPS, and to investigate MIPS, ${\bm \xi}^t_i(t)$ can be neglected~\cite{fily2012athermal}.

\begin{figure}[htbp]
\centering
{\includegraphics[width=1.\linewidth]{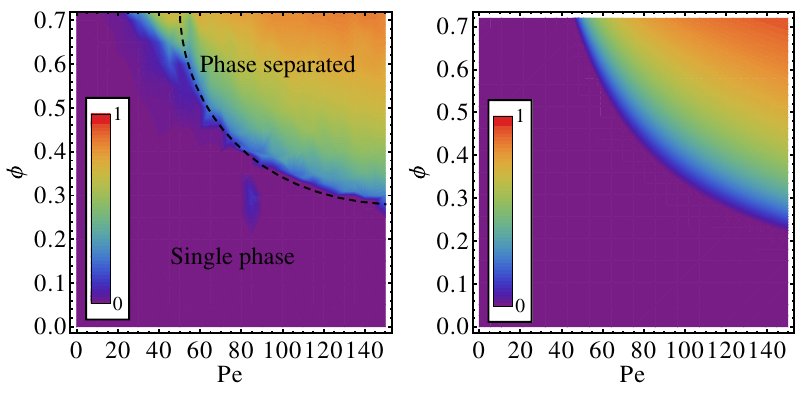}}
\caption{\label{fig2o}
Left: contour map of cluster fraction $f_c(\mathrm{Pe},\phi)$ measured from simulations. The dashed curve marks the approximate location of the binodal. Right: cluster fraction as predicted by the kinetic theory (Eq.~\ref{k_theory})~\cite{redner2013structure}.
Reproduced with permission.\cite{redner2013structure} Copyright 2013, American Physical Society.
}
\end{figure}

To understand the intriguing phase separation in ABPs, Ref.~\cite{redner2013structure} offers a kinetic theory to describe the steady state coexistence of dilute and dense phases, in which the dense phase is assumed to be close-packed. The orientations of particles in the dense phase evolve diffusively, while their positions are stationary. The dilute phase is treated as a homogeneous isotropic gas of density $\rho_g$, and if an ABP in the dilute phase collides with the dense phase, it gets absorbed immediately. Then one can write down the absorption rate of the particle with orientation $\theta$ with respect to the normal of dense phase surface as $k_{in}(\theta) = \frac{1}{2\pi} \rho_g F_p  \gamma_t^{-1} \cos(\theta)$, which leads to the total incoming flux per unit length: $k_{in} = \frac{\rho_g F_p}{\pi \gamma_t}$. On the other hand, the evaporation rate of ABPs $k_{out}$ from the dense phase is proportional to the rotational diffusion coefficient of ABPs $D_r$, and it can be written as $k_{out} = \frac{\kappa D_r}{\sigma}$ with a fitting parameter $\kappa$. In a steady-state coexistence, $k_{in} = k_{out}$, which leads to the prediction of fraction of particles in the dense phase:
\begin{equation}\label{k_theory}
f_c = \frac{4\phi \mathrm{Pe} - 3\pi^2 \kappa}{4 \phi \mathrm{Pe} - 6 \sqrt{3} \pi \kappa \phi},
\end{equation}
where $\phi = N\pi \sigma^2/(4V)$ is the packing fraction of the coexisting system with $N$ and $V$ the number of ABPs and the volume of the system, respectively. The comparison of the theoretical prediction (Eq.~\ref{k_theory}) with the results measured in Brownian dynamics simulations are shown in Fig.~\ref{fig2o}, in which a quantitative agreement can be found.

\subsection{Equilibrium-like mean field theory}
Besides the kinetic theory, one of the major interests for MIPS in the past years has been developing an equilibrium-like mean field theory to explain MIPS. In equilibrium, the characterization of phase separation includes the boundary of binodal and spinodal, and the kinetic of coarsening and nucleation. The free energy of an equilibrium homogeneous state can be written as a function of density $F(\rho) = \int f(\rho) \dd \mathbf{r}$ with $f(\rho)$ the free energy density, and the phase separation requires $F(\rho)$ to have a concave shape. As the coexisting phases have the same pressure and chemical potential, the binodal boundary is determined by the common tangent construction on $f(\rho)$ with respect to $\rho$.
The spinodal is obtained by the perturbation instability $\partial^2 f/\partial \rho^2<0$. From the kinetic perspective, the Cahn-Hillard theory predicts a power law growth of the domain size $L(t) \sim t^{\alpha}$, and if the order parameter $c_q$ indicating the domain follows the diffusive transport of chemical potential $\mu$, i.e., $\partial c_q/\partial t=D_q\nabla^2 \mu$ with $D_q$ the diffusion coefficient, $\alpha=1/3$ is obtained~\cite{bray2002theory}.
Using the concepts in equilibrium phase separations, various methods have been developed to construct an effective free energy for ABP systems, with a special focus on the MIPS in repulsive ABPs.

\subsubsection{ {Local density dependent effective bulk free energy}}
\label{SlocalV}
\begin{figure}[htbp]
\centering
{\includegraphics[width=1.\linewidth]{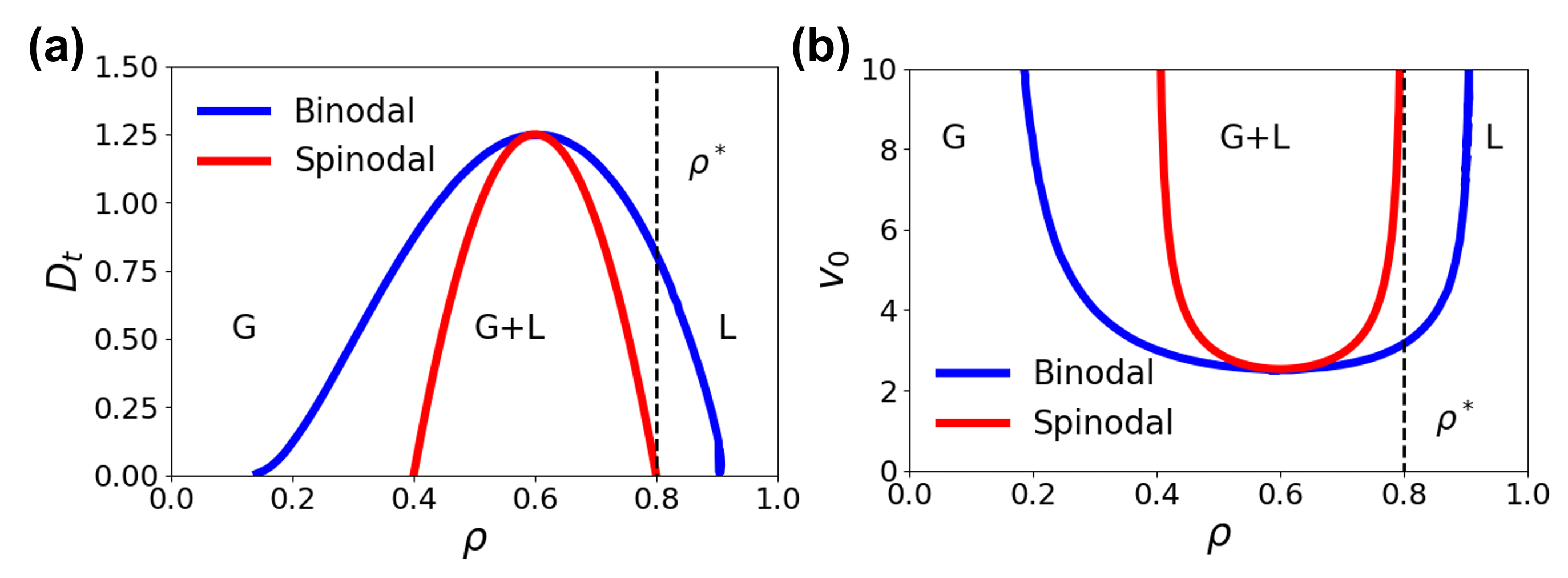}}
\caption{\label{fig2} Phase diagrams of MIPS for $v(\rho)$ with (a) $v_0=5$ and (b) $D_t=0.5, d=2, \tau=1, \rho^*=0.8$ ~\cite{cates2015motility}.
}
\end{figure}

In Ref.~\cite{cates2015motility}, by coarsening the zeroth harmonic of dynamic $\psi(\mathbf{r},\mathbf{u},t)$ for individual particles, which is the probability of finding a particle at position $\mathbf{r}$ moving in the direction $\mathbf{u}$ at time $t$, the many body Langevin equation $\rho(\mathbf{r},t)$ in the drift-diffusion form can be written as:
\begin{eqnarray}
\begin{aligned}
	\dot{\rho} &=\nabla \cdot (\mathbf{V}[\rho]\rho -D[\rho]\nabla \rho +(2D\rho)^{1/2} \mathbf{\Lambda}),\\
	\mathbf{V} &=\frac{-v\nabla v}{d(d-1)D_r},\\
	D &=\frac{v^2}{d(d-1)D_r}+D_t,
\end{aligned}
\end{eqnarray}
where the functionals $v[\rho]$ and $D_{t,r}[\rho]$ of the individual particle define the many-body drift velocity $\mathbf{V}[\rho]$ and diffusivity $D[\rho]$ for the interacting particle system. Here $d$ is the dimensionality of the system, and $D_t$ is the translational diffusional constant induced by the thermal noise in the dilute limit, i.e., ${\bm \xi}^t_i(t)$. $\mathbf{\Lambda}$ is a vector-valued unit white noise.
For this drift-diffusion system, an effective free energy functional can be introduced for the steady state as:
\begin{eqnarray}
\label{eqFfunctional}
	F[\rho]=F_{ex}[\rho]+\int \dd \mathbf{r} \rho (\ln{\rho}-1),
\end{eqnarray}
where $F_{ex}[\rho]$ is the effective excess free energy satisfying
\begin{eqnarray}
\label{eqVD1}
	\frac{\mathbf{V}([\rho],\mathbf{r})}{D([\rho],\mathbf{r})}=-\nabla \frac{\delta
	F_{ex}[\rho]}{\delta \rho(\mathbf{r})}.
\end{eqnarray}
With a local density-dependent velocity assumption, i.e., $v[\rho]=v(\rho)$, the local free energy density can be written as
\begin{eqnarray}
\label{bulkf1}
	f=\int_{0}^{\rho} \dd s \frac{1}{2} \ln \left[v^{2}(s) \tau+d D_{t}\right]+\rho(\ln \rho-1).
\end{eqnarray}
For systems of ABPs, it has been found that the local velocity almost linearly decreases with increasing density, i.e., $v(\rho)=v_0(1-\rho/\rho^*)$ with $v_0$ the velocity of a single isolated active particle, and $\rho^*$ the nearly close-packed density of the system (Ref.~\cite{fily2012athermal, redner2013structure, stenhammar2013continuum, stenhammar2014phase}), and the predicted phase boundary is shown in Fig.~\ref{fig2}.

\subsubsection{Beyond the effective bulk free energy}
\begin{figure}[htbp]
\centering
{\includegraphics[width=1.\linewidth]{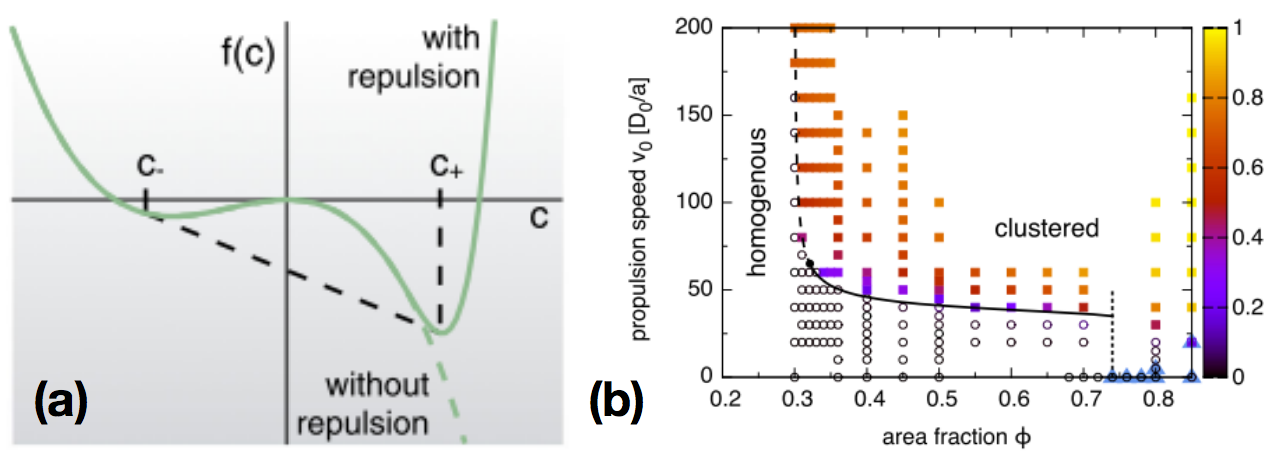}}
\caption{\label{fign4} (a) Concave shape of bulk effective free energy $f(c)$ for $\sigma_1<0$; (b) Spinodal boundary of MIPS, and the colorbar indicates the fraction of particles that are part of the largest cluster~\cite{speck2014effective}.
Reproduced with permission.\cite{speck2014effective} Copyright 2014, American Physical Society.
}
\end{figure}

Ref.~\cite{speck2014effective} studied the evolution of an initial perturbation for propulsion speeds $v_0 = v_c (1+\varepsilon)$ in the vicinity of the linear stability limit $v_c$. 
By expanding density fluctuation as $\delta \rho = \varepsilon c + \varepsilon^2 c^{(2)} + ...$, the lowest order evolution $\dot{c}$ and an effective free energy $F[c]$ can be obtained as
\begin{eqnarray}
\label{bulkf2}
\begin{aligned}
	\partial_{t} c &=\nabla^{2}\left[\sigma_{1} c-g c^{2}+\kappa_1 c^{3}-\nabla^{2} c\right]=\nabla^{2} \frac{\delta F}{\delta c},\\
	F[c] &=\int \dd \mathbf{r}\left[\frac{1}{2}|\nabla c|^{2}+f(c)\right],\\
	f(c) &=\frac{1}{2} \sigma_{1} c^{2}-\frac{1}{3} g c^{3}+\frac{\kappa_1}{4} c^{4},
\end{aligned}
\end{eqnarray}
where $\sigma_1$, $g$ and $\kappa_1$ are the coefficients depending on density with the assumption of effective velocity $v(\rho)=v_0-\zeta \rho$ where $\zeta$ is a constant. As $f(c)$ is concave, a perturbation of propulsion velocity $v_0$ around the instability boundary induces a phase separation into densities $\rho + \varepsilon c_{-}$ and $\rho + \varepsilon c_{+}$ as shown in Fig.~\ref{fign4}a. Perturbation analysis leads to the Clausius-Clapeyron equation
\begin{eqnarray}
\label{CCeq}
	\frac{\dd \phi}{\dd v}=\frac{\phi}{v}c_{-}(\phi,v).
\end{eqnarray}

Approximating $c_-$ by the inflection point $\dd f^2(c)/\dd c^2|_{c_-}=0$, one can obtain a spinodal boundary in the good agreement with simulation results as shown in Fig.~\ref{fign4}b.

\begin{figure}[htbp]
\centering
{\includegraphics[width=1.\linewidth]{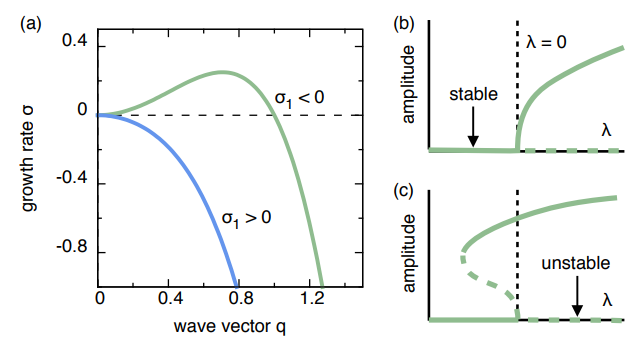}}
\caption{\label{fign5} (a) Growth rate as a function of the wave vector $q$ for the linearly unstable and linearly stable regimes; (b) Bifurcation diagram for the supercritical and (c) the subcritical case, where $\lambda$ is defined by $q^2=q_0^2-\lambda \varepsilon^2$, and $q_0$ is the instability boundary wavemode in dispersion relation~\cite{speck2014effective}.
Reproduced with permission.\cite{speck2014effective} Copyright 2014, American Physical Society.
}
\end{figure}

Besides the spinodals being well predicted from the bulk free energy density $f(c)$, the bifurcation analysis obtains the correct phase transition type, i.e.,
 a continuous transition at high density and a discontinuous transition at low density, which are usually described as spinodal decomposition and nucleation and growth scenarios, respectively. This can be seen in Eq.~\ref{bulkf2}, where the amplitude $a(q)$ of fluctuation $c(\mathbf{r})=a\exp [i\mathbf{q}\cdot \mathbf{r}]+c.c.$ shows two types of bifurcations
at the boundary of linear stable wave mode $q_0$ as shown in Fig.~\ref{fign5}, and the supercritical bifurcation leads to a continuous transition, while the subcritical bifurcation results in a discontinuous transition.

\subsubsection{Nonintegrable gradient term}
In Eq.~\ref{bulkf1}, by neglecting the contribution of $D_t$, as generally it is tiny compared with the propulsion term, the chemical potential can written as
\begin{eqnarray}
\label{EqChemicalP}
	\mu_0 \equiv \frac{\delta F[\rho]}{\delta \rho} =\ln \rho +\ln v[\rho].
\end{eqnarray}
In Ref.~\cite{stenhammar2013continuum}, beyond the local $v(\rho)$ assumption made in Sec.~\ref{SlocalV}, a gradient term is introduced to the velocity functional $v(\hat{\rho})$, where $\hat{\rho}(\mathbf{r})=\rho+\gamma^2\nabla^2 \rho$, so that it samples $\rho$ on a non-local length scale $\gamma(\rho)=\gamma_0 \tau_r v(\rho)$ depending on the orientational relaxation time $\tau_r$ with the coefficient $\gamma_0$. Substituting $v(\hat{\rho})$ into Eq.~\ref{EqChemicalP}, one can obtain a modified chemical potential
\begin{eqnarray}
\label{EqVDBmu}
\begin{aligned}
	\mu &=\mu_{0}-\kappa(\rho) \nabla^{2} \rho+\mathcal{O}\left(\nabla^{4} \rho\right), \\
	\kappa(\rho) &=-\gamma_{0}^{2} \tau_{r}^{2} v(\rho) \frac{\dd v(\rho)}{\dd \rho} .
\end{aligned}
\end{eqnarray}
However, such modification of the chemical potential breaks the detailed balance (DB), as it can not be integrated into any effective free energy functional. Meanwhile, using similar ideas, one can introduce the density dependent surface tension parameter $\kappa (\rho)$ directly into the free energy density and obtain the corresponding chemical potential recovering DB
\begin{eqnarray}
\label{EqRDBmu}
\begin{aligned}
	f[\rho] &=f_{0}(\rho)+\frac{\kappa(\rho)}{2}(\nabla \rho)^{2},\\
	\mu &=\mu_{0}-\kappa(\rho) \nabla^{2} \rho-\frac{\dd \kappa(\rho)}{\dd \rho} \frac{(\nabla \rho)^{2}}{2}.
\end{aligned}
\end{eqnarray}
Then the corresponding continuum model can be written as
\begin{eqnarray}
\label{EqCMmu}
	\partial_{t} {\phi}=\nabla \cdot\left\{{\phi}(1-{\phi})^{2} \nabla \mu-\sqrt{2 {\phi}(1-{\phi})^{2} N_{0}^{-1}} \mathbf{\Lambda}\right\},
\end{eqnarray}
where $\phi(\mathbf{r}) \sim \rho(\mathbf{r})$ is the local packing fraction and $v({\phi}) = v_0 (1-{\phi})$~\cite{stenhammar2013continuum}. Here $N_0$ is the number of ABPs in the cell of side $D_t/v_0$.

\begin{figure}[htbp]
\centering
{\includegraphics[width=1.\linewidth]{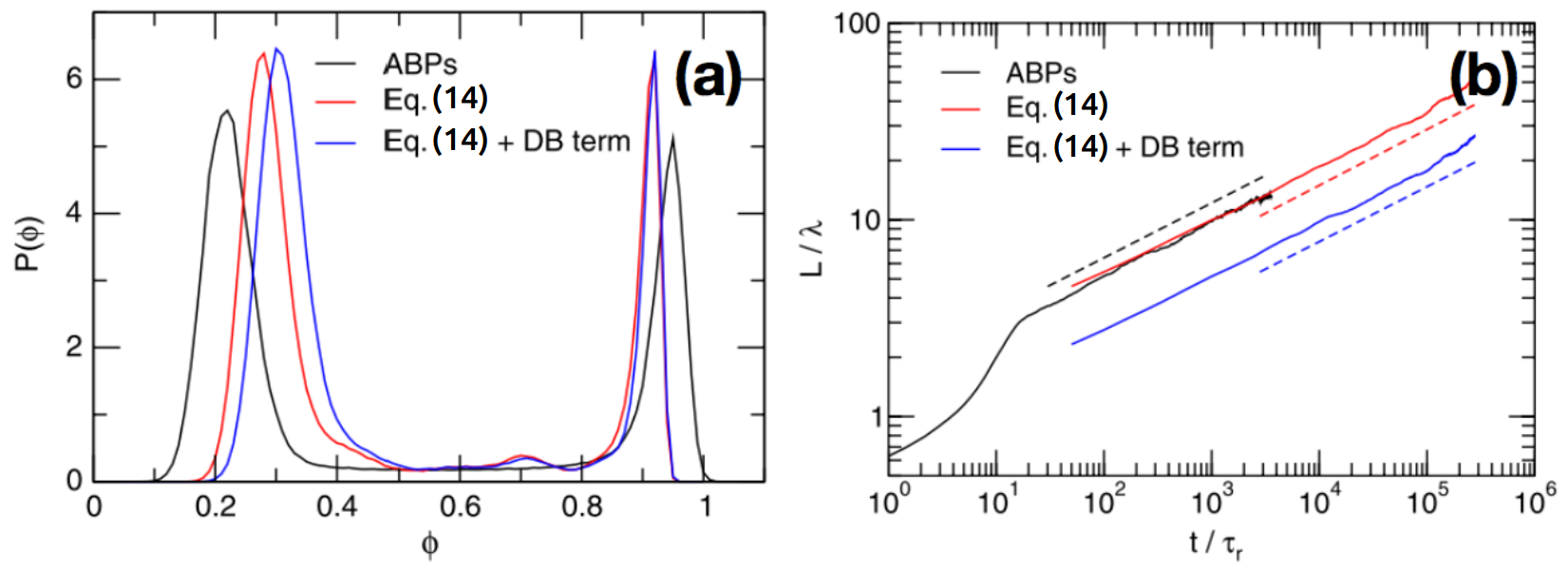}}
\caption{\label{fignn5} (a) Probability distribution $P(\phi)$ of the local area fraction $\phi$; (b) Time-dependent coarsening length $L(t) \sim t^\alpha$, where the fitted exponents are $\alpha=0.27(9)$ (ABPs), $\alpha=0.28(7)$ (continuum model without DB), $\alpha=0.27(9)$ (continuum model with DB)~\cite{stenhammar2013continuum}.
Reproduced with permission.\cite{stenhammar2013continuum} Copyright 2013, American Physical Society.
}
\end{figure}

The comparison between computer simulations of ABPs and the numerical solution of the continuum models (Eq.~\ref{EqCMmu}) with (Eq.~\ref{EqRDBmu}) and without DB (Eq.~\ref{EqVDBmu}) is shown in Fig.~\ref{fignn5}, including both the obtained binodal (Fig.~\ref{fignn5}a) and the coarsening dynamics (Fig.~\ref{fignn5}b). Here the coarsening length scale is defined as $L(t)=2 \pi \frac{\int S(k, t) \dd k}{\int k S(k, t) \dd k}$, where $S(k, t)$ is the structure factor of the system at time $t$.
Interestingly, even though the DB violation enters this model via the gradient term in the effective free energy, which is generally related with the interface formation, it has little influence on the coarsening dynamics, while indeed changes the coexisting binodal.

\subsection{Binodal of MIPS\label{secbinodal}}
In the previous section, we reviewed the progress in mapping the dynamics of ABPs to equilibrium systems via $\dot{\rho} \sim \nabla \cdot (\nabla \frac{\delta {F}}{\delta \rho})$ \cite{cates2015motility}, which has obtained a good estimation of the spinodal, phase transition types,  and coarsening dynamics \cite{speck2014effective, stenhammar2013continuum} for MIPS.
However, as shown in Fig.~\ref{fignn5}a, the predicted binodal is still significantly different from the measurements in computer simulations~\cite{stenhammar2013continuum}, and the thermodynamic pressure $p=\partial F/ \partial V$ was found unequal across the interface between coexisting phases ~\cite{wittkowski2014scalar}. These violation suggests that the steady state of ABPs does not correspond to the extrema of thermodynamic free energy ${F}[\rho]$. This is not surprising as part of the free energy budget is dissipated, and the detailed balance is broken.

In Ref.~\cite{solon2018generalized, solon2018generalizedR}, by defining a one-to-one mapping $R(\rho)$, generalized thermodynamic principles are developed for the dynamics $\dot{\rho} \sim \nabla \cdot (\nabla \frac{\delta \mathcal{G}}{\delta R})$, where $\mathcal{G}=\int \dd \mathbf{r} \left(\Phi(R)+\Phi_1[R] \right)$ is the generalized free energy with $\Phi(R)$ and $\Phi_1[R]$ free energy densities.
Then the violations mentioned above are reasonable as the steady state is at the extrema of $\mathcal{G}$ instead of $F$.
Like in equilibrium, two intensive quantities are found equal in coexisting phases of MIPS, namely the generalized chemical potential $g_0(R)=\frac{\dd \Phi}{\dd R}$ and the generalized pressure $h_0(R)=R\frac{\dd \Phi}{\dd R}-\Phi$, so that the common tangent of $g_0(R)$ and the equal-area Maxwell construction of $h_0(1/R)$ can be employed to determine the binodal of MIPS.

To apply this generalized thermodynamics into ABPs, the hydrodynamic description of conserved density $\partial_t \rho(\mathbf{r},t)=-\nabla \cdot \mathbf{J}$ is obtained by integrating the probability density distribution function of particle $\psi(\mathbf{r},\theta,t)$ at position $\mathbf{r}$ with orientation $\theta$, and
\begin{eqnarray}
\label{EQHydroABP}
	\mathbf{J}=v_0\mathbf{m}+\mathbf{I}^{(0)}-D_t\nabla \rho,
\end{eqnarray}
with $\mathbf{m}(\mathbf{r})=\int d\theta ~\mathbf{u} \psi$ the polarization field, which is the first harmonic of $\psi$ with $\mathbf{u} = (\cos \theta, \sin \theta)$, and $\mathbf{I}^{(0)}(\mathbf{r})=-\int \operatorname{d\mathbf{r}}^{\prime} u_t \nabla V\left(\left|\mathbf{r}-\mathbf{r}^{\prime}\right|\right)\left\langle {\rho}(\mathbf{r}) {\rho}\left(\mathbf{r}^{\prime}\right)\right\rangle$ the pairwise interaction density, where $u_t=1/\gamma_t$ is the mobility, and $V(\cdot)$ is the pair potential between ABPs. $\dot{\mathbf{m}}$ depends on the second harmonic $\mathbb{Q}=\int \dd \theta (\mathbf{u} \colon \mathbf{u}-\mathbb{I}/2)\psi$ (nematic order field), and $\dot{\mathbb{Q}}$ depends on the higher order harmonic $\Theta[\psi]$. To close Eq.~\ref{EQHydroABP}, one can make the quasi-stationary assumption $\dot{\mathbf{m}}=\dot{\mathbb{Q}}=\dot{\Theta}=0$, since they are fast modes compared with $\rho(\mathbf{r},t)$~\cite{solon2018generalizedR, cates2013active}.

For the expression of flux in Eq.~\ref{EQHydroABP}, instead of constructing a $\mathbf{J} \sim \nabla \frac{\delta \mathcal{G}}{\delta R}$ and predicting the binodal using the first principle approach, in Ref.~\cite{kruger2018stresses} a stress tensor $\bm{\sigma}$ is found so that  $\mathbf{J} \sim \nabla \cdot \bm{\sigma}$, and the generalized pressure is defined as the diagonal term
\begin{eqnarray}
\label{EQGenerPressure}
\begin{aligned}
	h &\equiv-\sigma_{x x}\\
	&=\frac{D_{t}}{u_t} \rho+P^{A}(x)+P^{D}(x)+\frac{v_{0}^{2}}{u_t D_{r}} \mathbb{Q}_{x x}-\frac{D_{t} v_{0}}{u_t D_{r}} \partial_{x} m_{x}.
\end{aligned}
\end{eqnarray}
Here $P^A$ and $P^D$ are ``active'' and ``direct'' passive-like part pressures, respectively, which have explicit mechanical definitions~\cite{solon2015pressure} discussed in detail in Sec.~\ref{SECPressure}.

\begin{figure}[!ht]
\centering
\includegraphics[width=1.\linewidth]{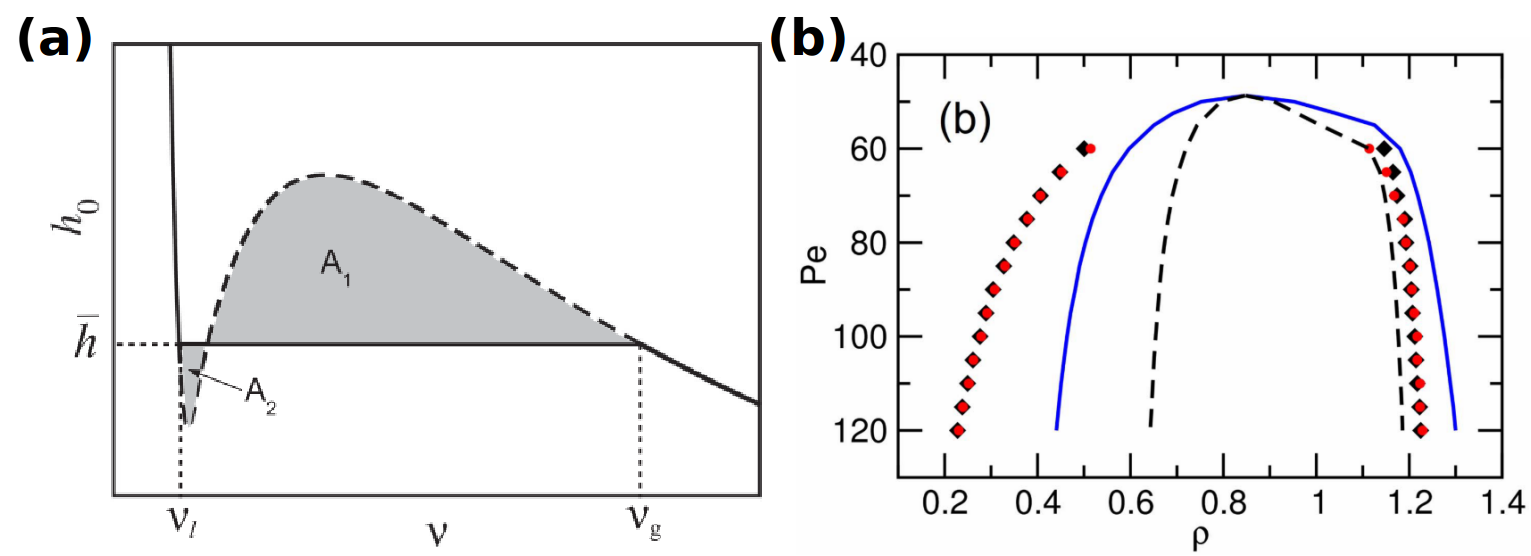}
\caption{\label{FigBinodalABP} (a) Modified Maxwell construction: for EOS $h_0(\nu)$, $h_0(\nu_{\ell})=h_0(\nu_{g})$ and $A_1-A_2=\Delta A$ are used to determine $\nu_{\ell,g}$. (b) Phase diagram of ABPs, modified(red) and the equal-area (blue) Maxwell constructions, compared with numerically measured binodals (black). Dashed lines correspond to the spinodals $h_0'(\rho)=0$~\cite{solon2018generalizedR}.
Reproduced with permission.\cite{solon2018generalizedR} Copyright 2018, IOP Publishing.
}
\end{figure}

The flux-free steady state sets the first restriction to coexisting homogeneous phases
\begin{eqnarray}
	h_0(\rho_{\ell})=h_0(\rho_g)=\bar{h},
\end{eqnarray}
where $\bar{h}$ is the coexisting pressure. The generalized pressure $h$ is splitted into a local term and an interfacial contribution: $h=h_0(\rho)+h_1[\rho]$. To predict binodals, we need one more restriction, as no explicit $R(\rho)$ is obtained, and the equal-area Maxwell construction for the equation of state (EOS) $h_0(1/R)$ cannot be applied. Instead, Solon \emph{et al.} measured the violation of Maxwell construction for $h_0(1/\rho)$ numerically in a slab simulation box~\cite{solon2018generalizedR}
\begin{eqnarray}
\label{EQmodMC}
	\int_{\nu_{\ell}}^{\nu_{g}}\left(h_{0}(\nu)-\bar{h}\right) \mathrm{d} \nu=\int_{x_{g}}^{x_{\ell}} h_{1} \partial_{x} \nu \mathrm{d} x \equiv \Delta A,
\end{eqnarray}
where $\nu=1/\rho$. The integration in Eq.~\ref{EQmodMC} goes through the interface separating the coexisting gas and liquid-like phases along the $x$ axis of the simulation box.
Then for a given $\Delta A$, coexisting phases were determined on the EOS $h_0(\nu)$ as schematic in Fig.~\ref{FigBinodalABP}a, and the phase diagram obtained fit well with simulation result Fig.~\ref{FigBinodalABP}b.

\subsection{Pressure}
\label{SECPressure}
The generalized pressure defined in Eq.~\ref{EQGenerPressure} coincides with the mechanical pressure in the isotropic homogeneous state of ABPs \cite{solon2015pressure}.

\begin{figure}[!ht]
\centering
\includegraphics[width=1.\linewidth]{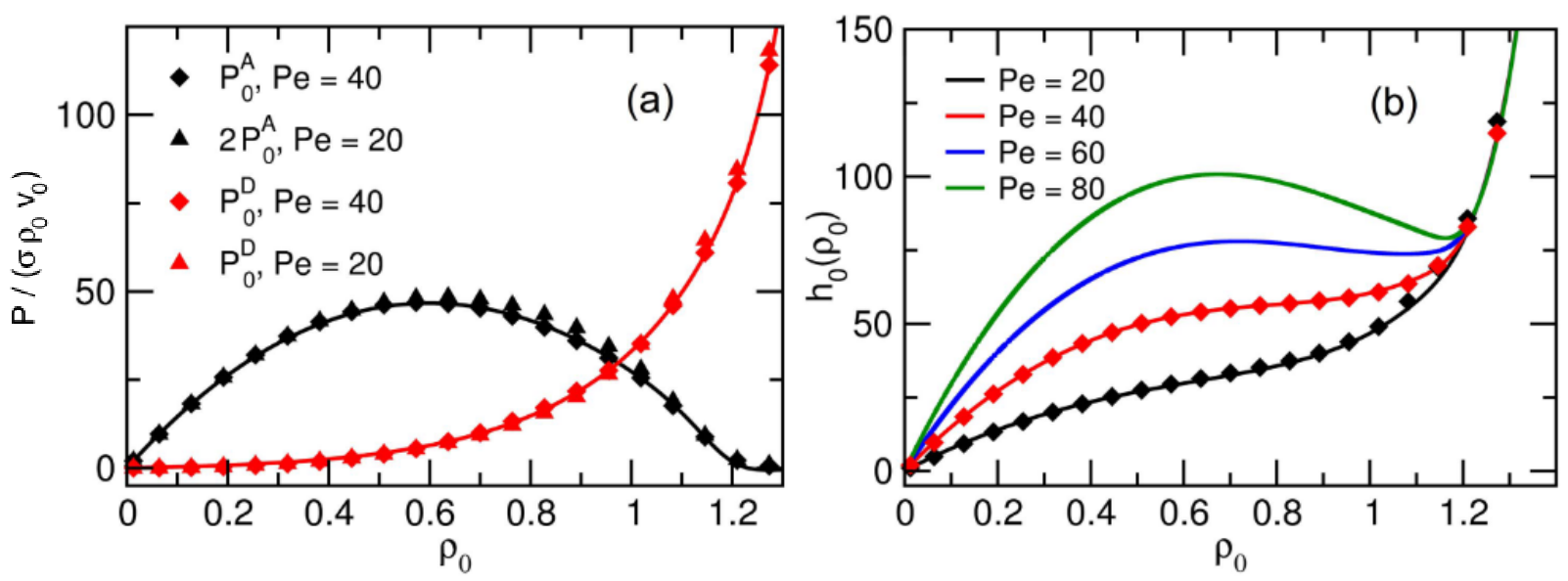}
\caption{\label{FigNumericalEOS} (a) $P_0^A$ (black) scales linearly with Pe and $P_0^D$ (red) is independent with Pe. (b) Empirical EOS (lines) $h_0(\rho)=P_0^A+P_0^D$ constructed by scale rules of Pe based on numerically measured EOS (dots) at homogeneous state~\cite{solon2018generalizedR}.
Reproduced with permission.\cite{solon2018generalizedR} Copyright 2018, IOP Publishing.
}
\end{figure}

The ``active'' contribution to pressure is
\begin{eqnarray}
	P^A=\frac{v_0^2}{2u_t D_r}\rho +\frac{v_0}{u_t D_r}\mathbb{I}_{xx}^{(1)},
\end{eqnarray}
which is also called ``swim'' pressure proposed by Brady and coworkers~\cite{takatori2014swim, takatori2015towards}, and
\begin{eqnarray}
	\mathbb{I}_{\alpha \beta}^{(1)}=-\int \mathrm{d} \mathbf{r}^{\prime} u_t \partial_{\beta} V\left(\left|\mathbf{r}-\mathbf{r}^{\prime}\right|\right)\left\langle \rho \left(\mathbf{r}^{\prime}\right) m_{\alpha}(\mathbf{r})\right\rangle
\end{eqnarray}
describes the pairwise interaction density projected on the self-propulsion direction. Therefore, the physical interpretation of $P^A$ is the transport of propulsion forces~\cite{solon2015pressure}. Assuming short-range repulsions between ABPs, $P^A$ can be estimated as $P^A=v_0v(\rho)\rho/(2u_t D_r)$ with $v(\rho)=v_0(1-\rho/\rho^*)$ \cite{fily2012athermal, redner2013structure, stenhammar2013continuum}.
When neglecting the contribution of $D_t$, Solon \emph{et al.} defined a P{\'{e}}clet number as $\mathrm{Pe} = 3v_0/D_r\sigma$~\cite{solon2018generalizedR}, and it was found that at small Pe regime, i.e., $<\text{Pe}_c$, the system remains in a homogeneous state.
Here $\text{Pe}_c$ is the critical P{\'{e}}clet number, above which MIPS occurs.
For fixed $v_0$, by changing $D_r$, $P_0^{A} \propto \mathrm{Pe}$, which was numerically confirmed in Fig.~\ref{FigNumericalEOS}a, where the subscription 0 means in a homogeneous state.

The ``direct'' contribution to the pressure
\begin{equation}
	P^D= -\sigma_{xx}^{\text{IK}},
\end{equation}
with
\begin{eqnarray}
\begin{aligned}
	\sigma_{\alpha \beta}^{\mathrm{IK}}(\mathbf{r})= &\frac{1}{2} \int \mathrm{d} \mathbf{r}^{\prime} \frac{r_{\alpha}^{\prime} r_{\beta}^{\prime}}{\left|\mathbf{r}^{\prime}\right|} \frac{\mathrm{d} V\left(\left|\mathbf{r}^{\prime}\right|\right)}{\mathrm{d}\left|\mathbf{r}^{\prime}\right|}\\
	&\int_{0}^{1} \mathrm{d} \lambda\left\langle \rho \left(\mathbf{r}+(1-\lambda) \mathbf{r}^{\prime}\right)  \rho \left(\mathbf{r}-\lambda \mathbf{r}^{\prime}\right)\right\rangle,
\end{aligned}
\end{eqnarray}
representing the density of pairwise forces acting across a plane~\cite{solon2015pressure}, and it is passive-like and independent with Pe in a homogeneous phase (Fig.~\ref{FigNumericalEOS}a). As $h_0(\rho)=P_0^A+P_0^D$, the EOS of phase separated state at high Pe can be constructed following the scale rule
\begin{equation}
h_0(\rho,\mathrm{Pe}) = P_0^{A}(\rho,\mathrm{Pe}_0)\left( \frac{\mathrm{Pe}}{\mathrm{Pe}_0}\right) + P_0^D(\rho,\mathrm{Pe}_0),
\end{equation}
where $\mathrm{Pe}_0 < \text{Pe}_c$ (Fig.~\ref{FigNumericalEOS}b).

\begin{figure}[!ht]
\centering
\includegraphics[width=1.\linewidth]{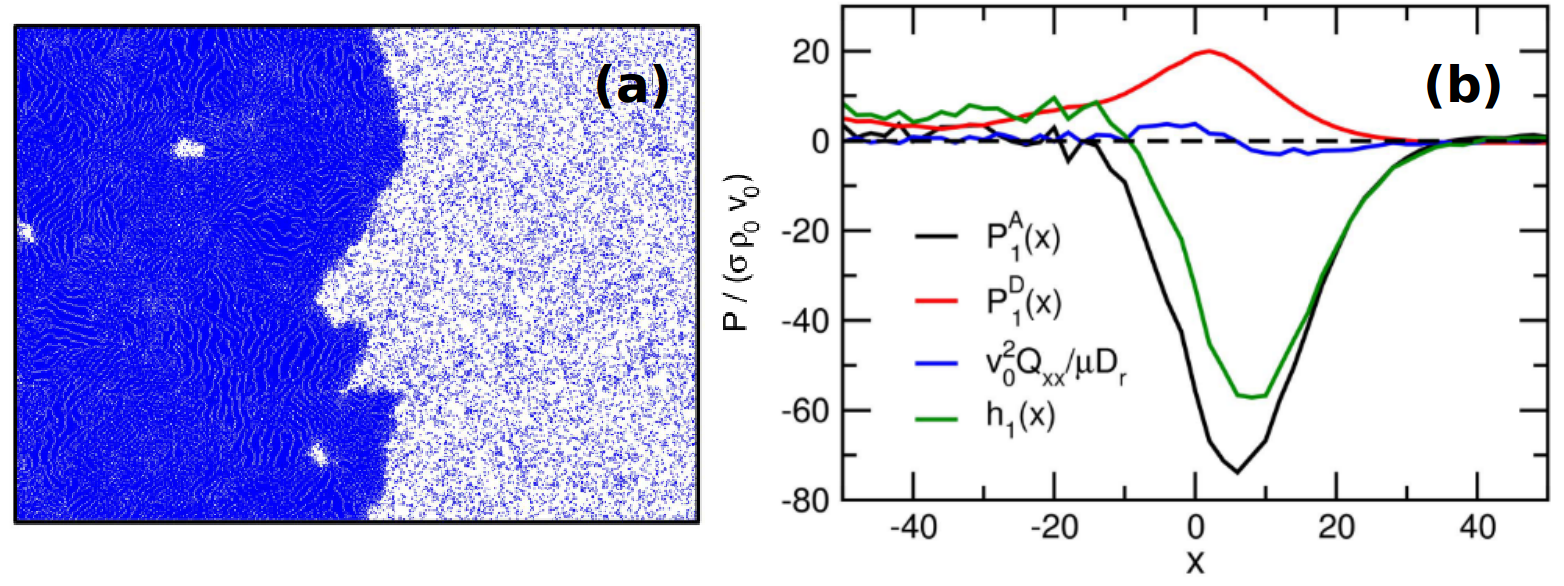}
\caption{\label{FigPressureInterface} (a) Flat interface between phase-separated state in slab simulation box. (b) Nonlocal terms contribute to generalized pressure at the flat interface~\cite{solon2018generalizedR}.
Reproduced with permission.\cite{solon2018generalizedR} Copyright 2018, IOP Publishing.
}
\end{figure}

The modified Maxwell construction in Eq.~\ref{EQmodMC} results from the fact that the integration of nonlocal contribution $h_1[\rho]=P_{1}^{A}[\rho]+P_{1}^{D}[\rho]+\frac{v_{0}^{2}}{u_t D_{r}} \mathbb{Q}_{x x}-\frac{D_{t} v_{0}}{u_t D_{r}} \partial_{x} m_{x}$ at interface does not vanish, where the subscription $1$ indicates the interfacial contribution, 
and the contribution of each term can be numerically measured for a flat interface in a slab simulation box (Fig.~\ref{FigPressureInterface}a). The nonlocal contribution of ``active'' pressure $P_1^A$ is negative, while the ``direct'' pressure $P_1^D$ is positive, which is due to the more intensive collision events occuring at the interface. Overall, this leads to a negative interfacial tension (Fig.~\ref{FigPressureInterface}b), which is also observed in Ref.~\cite{bialke2015negative}.

 {Moreover, apart from the derivation of a generalized EOS $h_0(R)$ from the generalized free energy $\mathcal{G}$ as discussed in Sec.~\ref{secbinodal}, the existence of an EOS in systems of torque-free ABPs was also confirmed via mechanical approaches~\cite{fily2017mechanical, das2019local}. It was found that for torque-free ABP systems, there is an effective conservation of momentum in the steady state, which leads to a mechanical pressure depending solely on bulk quantities~\cite{fily2017mechanical}.
A local stress expression for ABPs using the virial theorem derived in Ref.~\cite{das2019local} also underlines the existence of an EOS, and opens up the possibility to calculate stresses even in inhomogeneous systems.}

\section{Anisotropic active colloids}\label{AACSEC}
Different from the system of active spherical particles, in which MIPS was one of the major research focuses in the past, in systems of anisotropic self-propelled particles, the coupling between the intrinsic shape anisotropy and the self-propulsion results in effective alignments, which have produced more emergent dynamic phases like swarming, turbulent, lane and oscillation states.
In Sec.~\ref{SECstericSPR}, we summarize recent simulations on self-propelled rods (SPRs) interacting with steric repulsion in two-dimensions, and we also discuss systems of active particles with more complex shape in Sec.~\ref{SECChiral}, where the chirality plays an important role in assembly.
Further, in Sec.~\ref{SECVicsekLike}, we compare with point polar particle systems with explicit alignment interactions, which are more convenient for theoretical analysis, and various modified models are designed to bridge them with the SPR systems.
Finally, in Sec.~\ref{SECCharacter}, we summarize useful properties of density fields, orientational fields and swirling flows to characterize various phases and phase transitions.

\subsection{SPR model and phase diagrams}
\label{SECstericSPR}
Self-propelled rigid chains of repulsive disks are one of the most commonly used models for investigating the dynamic assembly of SPRs in two-dimensions. As shown in Fig.~
\ref{FigSPRModelDiagram}a, $\mathbf{x}_i$ and $\theta_i$ are the position of the center of mass and the orientation of the long axis of the rod, respectively. The equation of motion for the SPR $i$ is
\begin{eqnarray}
\begin{aligned}
	\dot{\mathbf{x}}_{i}=& \bm{\mu}\left[-\nabla U_{i}+F \mathbf{V}\left(\theta_{i}\right)+\mathbf{\Lambda}_{i}(t)\right], \\
	\dot{\theta}_{i} &=\frac{1}{\zeta_{\theta}}\left[-\frac{\partial U_{i}}{\partial \theta_{i}}+\xi_{i}(t)\right],
\end{aligned}
\end{eqnarray}
where $\bm{\mu}$ is the mobility tensor defined as $\zeta_{\|}^{-1} \mathbf{V} (\theta_{i}) \mathbf{V}(\theta_{i})+\zeta_{\perp}^{-1}(\mathbf{I}-\mathbf{V}(\theta_i) \mathbf{V}(\theta_i))$, with $\zeta_{\|,\perp,\theta}$ the friction coefficients and $\mathbf{V}(\theta_i) = (\cos \theta_i, \sin \theta_i)$.
Here the interaction between two SPRs is simplified as the summation of the pair interactions between all disks, and $U_{ij}=\sum_{\alpha,\beta}u_{i,j}^{\alpha,\beta}$, where $u_{i,j}^{\alpha,\beta}$ is the potential between disk $\alpha$ of the $i$th rod and disk $\beta$ of the $j$th rod. $\mathbf{\Lambda}$ and $\xi$ are the vectorial and scalar Gaussian white noises, respectively. The excluded volume interaction between discs induces a torque, which leads to an effective nematic alignment between SPRs as shown in Fig.~\ref{FigSPRModelDiagram}.
\subsubsection{Control parameters: aspect ratio $a$ and packing fraction ${\phi}$}
\begin{figure}[!ht]
\centering
\includegraphics[width=1.\linewidth]{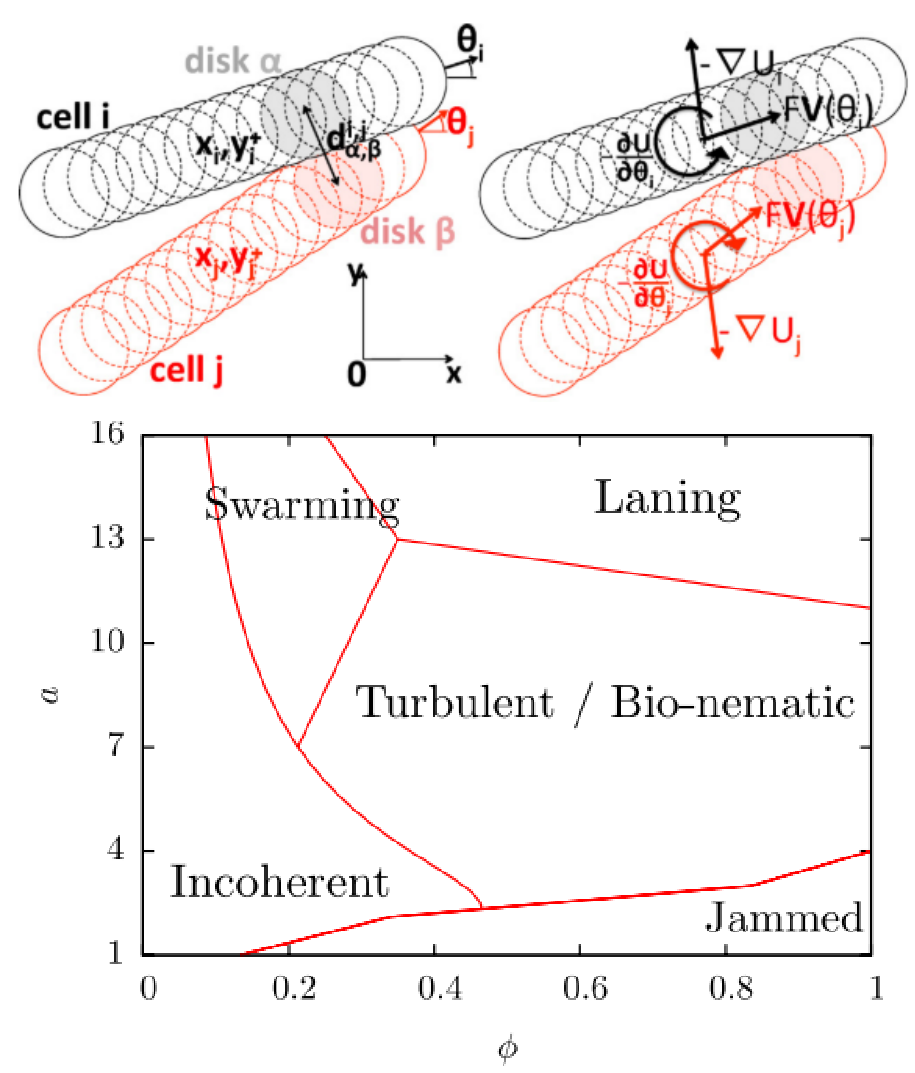}
\caption{\label{FigSPRModelDiagram} Top: Sketch of two interacting rods in the SPR model. Forces and torques induced by the interaction leading to the effective alignment (right panel)~\cite{weitz2015self}. 
Reproduced with permission.\cite{weitz2015self} Copyright 2015, American Physical Society.
Bottom: Schematic non-equilibrium phase diagram of the 2D SPR model at variable aspect ratio $a$ and effective filling fraction $\phi$ \cite{wensink2012emergent}.
Reproduced with permission.\cite{wensink2012emergent} Copyright 2012, IOP Publishing.
}
\end{figure}

Using the effective packing fraction $\phi = N [\sigma(l-\sigma) + \pi \sigma^2/4]/V$ and the length-to-width ratio $a = l/\sigma$ as control parameters, a phase diagram for SPRs is obtained as shown in Fig.~\ref{FigSPRModelDiagram} ~\cite{wensink2012meso, wensink2012emergent}, which features the following phases. Here $l$ and $\sigma$ are the length of rods and diameter of discs, respectively.

\textbf{Active jamming:} For small aspect ratio $a$, increasing $\phi$ results in a sharp drop of the mobility of SPRs, which marks the onset of jamming transition. The boundary highly depends on the shape anisotropy of SPRs.

\textbf{Turbulent:} For intermediate $a$, the significant increase of enstrophy density defined as $\Omega=\frac{1}{2}\langle |{w}(\mathbf{r},t)|^2 \rangle$ marks the transition to a turbulent phase, where particles collectively swirl.  {The scalar vorticity field is defined as $w(\mathbf{r},t)=[\nabla \times \mathbf{v}(\mathbf{r},t)]\cdot \hat{\mathbf{e}}_z$, where $\mathbf{v}(\mathbf{r},t)$ is the velocity field, and $\hat{\mathbf{e}}_z$ is the unit vector perpendicular to the 2D plane.}
We note that with increasing $\phi$, the system experiences a crossover from the state of inhomogeneous density distribution with giant number fluctuation (GNF) to a homogeneous turbulent state.

\textbf{Swarming:} In systems of slender SPRs, at low to moderate density, particles tend to form large compact flocks with coherent motion indicated by the relatively long velocity correlation length.

\textbf{Laning:} For extremely large $a$ at high density, the divergence of the correlation length with increasing $\phi$ marks the transition to the laning phase, where flocks start to span the entire system and self-organize into lanes moving in opposite directions.

\begin{figure}[!ht]
\centering
\includegraphics[width=1.\linewidth]{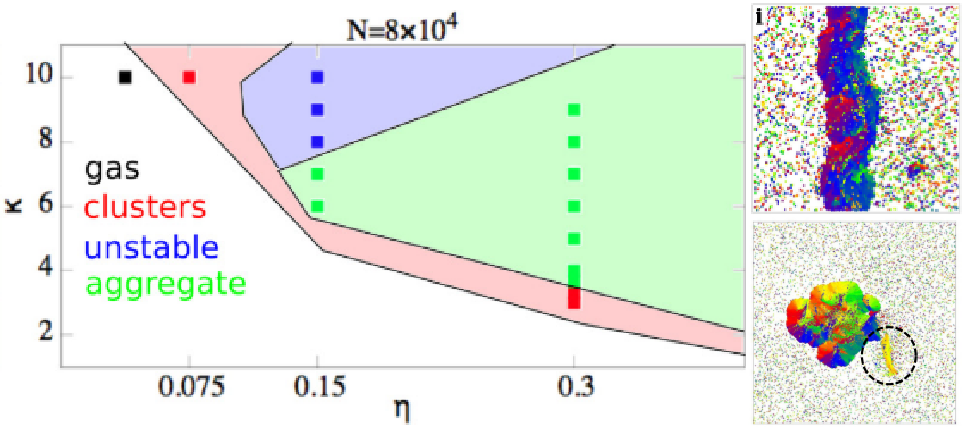}
\caption{\label{FigSPRClusterDiagram} Left: Phase diagram of the 2D disk-chain model at variable aspect ratio $\kappa$ and effective filling fraction $\eta$, where $\kappa$ and $\eta$ are the same as $a$ and $\phi$ in Fig.~\ref{FigSPRModelDiagram}. 
Right-top: Snapshots of a percolate polar cluster. Right-bottom: Giant aggregate state~\cite{weitz2015self}.
Reproduced with permission.\cite{weitz2015self} Copyright 2015, American Physical Society.
}
\end{figure}

The results above provide a schematic phase diagram of SPRs, while in another similar disc-chains model, an ``aggregate'' phase was found~\cite{weitz2015self}, of which the phase diagram is shown in Fig.~\ref{FigSPRClusterDiagram}.
The ``clusters'' phase in Fig.~\ref{FigSPRClusterDiagram} is similar to the ``swarming'' phase in Fig.~\ref{FigSPRModelDiagram}, while slightly increasing density makes those flocks with local polar order aggregate into a giant cluster, of which the size increases linearly with the system size and the global orientation order vanishes.
For larger aspect ratio, there is a broad ``unstable'' regime of density, where the system oscillates between clusters and giant aggregates states.
Comparing the difference between the two models, even though the former employs the Yukawa potential, which is steeper than the harmonic potential used in the latter model, they both achieve an effective velocity alignment in pairwise rod collision events. The difference is that the SPR model in Fig.~\ref{FigSPRModelDiagram} neglects both the translational and rotational thermal noises so that it is deterministic but the disc-chain model in Fig.~\ref{FigSPRClusterDiagram} considers the rotational thermal noise. Besides, the SPR model employs a mobility tensor depending on aspect ratio, while $\zeta_{\|,\perp,\theta}$ is fixed in the disc-chain model, and in the investigated regime of aspect ratio, the mobility of SPRs in the former model (Fig.~\ref{FigSPRModelDiagram}) is much smaller than that in the latter (Fig.~\ref{FigSPRClusterDiagram}). The most significant difference may be that the the self-propulsion velocity along long-axis in the SPR model investigated in Fig.~\ref{FigSPRModelDiagram} is much larger than the disc-chain in Fig.~\ref{FigSPRClusterDiagram}.
All these break the large swarming clusters in the SPR model into smaller polar clusters in the disc-chain model, so that the giant clusters formed in a ``traffic jammed'' scenario is possible in the disc-chain model.

\subsubsection{Control parameters: packing fraction $\phi$ and $\mathrm{Pe}$}
\begin{figure*}[!ht]
\centering
\includegraphics[width=1.\linewidth]{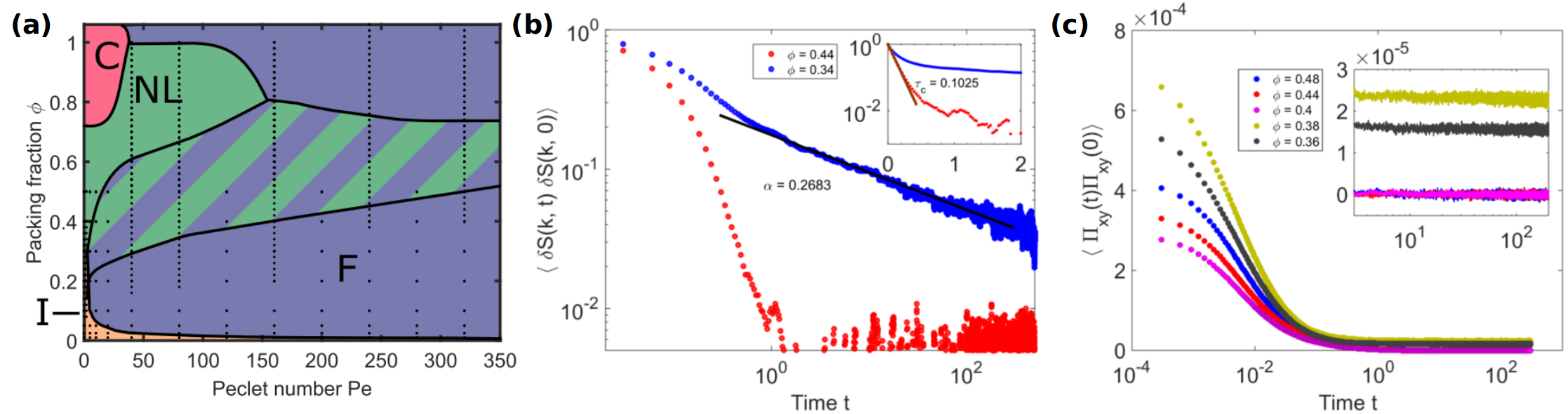}
\caption{\label{FigSPRHysteresisDiagram} (a) Phase diagram of the 2D SPR model with aspect ratio $40$ at variable packing fraction $\phi$ and Pe. (b) Structure factor peak correlation as a function of time for $\text{Pe}=160$ in the laning ($\phi=0.44$) and flocking ($\phi=0.34$) state. (c) Stress-tensor correlation as a function of time for $\text{Pe}=160$, inset, zoomed view of long-time tail~\cite{kuan2015hysteresis}.
Reproduced with permission.\cite{kuan2015hysteresis} Copyright 2015, American Physical Society.
}
\end{figure*}

Different from the studies above focusing on the control parameters aspect ratio $a$ and packing fraction $\phi$, in Ref.~\cite{kuan2015hysteresis}, the aspect ratio is fixed at $a=40$ and the phase behaviour of the system in the parameter space of $\phi, \text{Pe}$ was investigated in Fig.~\ref{FigSPRHysteresisDiagram}. At the passive limit $\text{Pe}=0$, with the increasing $\phi$, the system transforms from an isotropic fluid to a nematic and finally to a crystalline phase.
With increasing the self-propulsion, a flock phase appears, which is characterized by the collective motion of dense aligned clusters, indicated by the peaks in the pair distribution function and the emergence of the polar orientational correlation that persists over the length scale of typical cluster-size.  {Here the flock phase is essentially the same as the swarming phase defined in Ref.~\cite{wensink2012emergent}.}
The transition from a nematic phase to a flock phase has a strong hysteresis.

The normalized structure factor time correlation functions for nematic-laning and flocking phases are shown in Fig.~\ref{FigSPRHysteresisDiagram}b, and one can see that
the nematic-laning phase exhibits an exponential decay, while the flocking state shows a power-law decay, which indicates slow structural relaxation at the length scale of typical cluster-size in the flocking state. Besides, the long-lived non-zero plateau in the correlation function of the internal stress tensor suggests a slow mechanical relaxation. These imply that the transition from nematic-laning to flocking is glassy.

In experimental systems like actin filaments propelled by molecular motors, active units are penetrable, which significantly affects the phase behavior, as the motility is not restricted in two-dimensions. In Ref.~\cite{abkenar2013collective}, a separation-shifted Lennard-Jones potential is employed so that the SPRs have certain probability to penetrate each other. At the fixed aspect ratio $a=18$, for intermediate $\phi$, with increasing Pe, the system first forms a flocking phase similar to Fig.~\ref{FigSPRHysteresisDiagram}, while further increasing Pe breaks up the polar clusters and the system becomes homogeneous, indicated by the density probability distribution changing from unimodal to bimodal finally again to a broader unimodal distribution.

\subsection{Chirality and oscillation state}\label{SECChiral}
\begin{figure}[!ht]
\centering
\includegraphics[width=1.\linewidth]{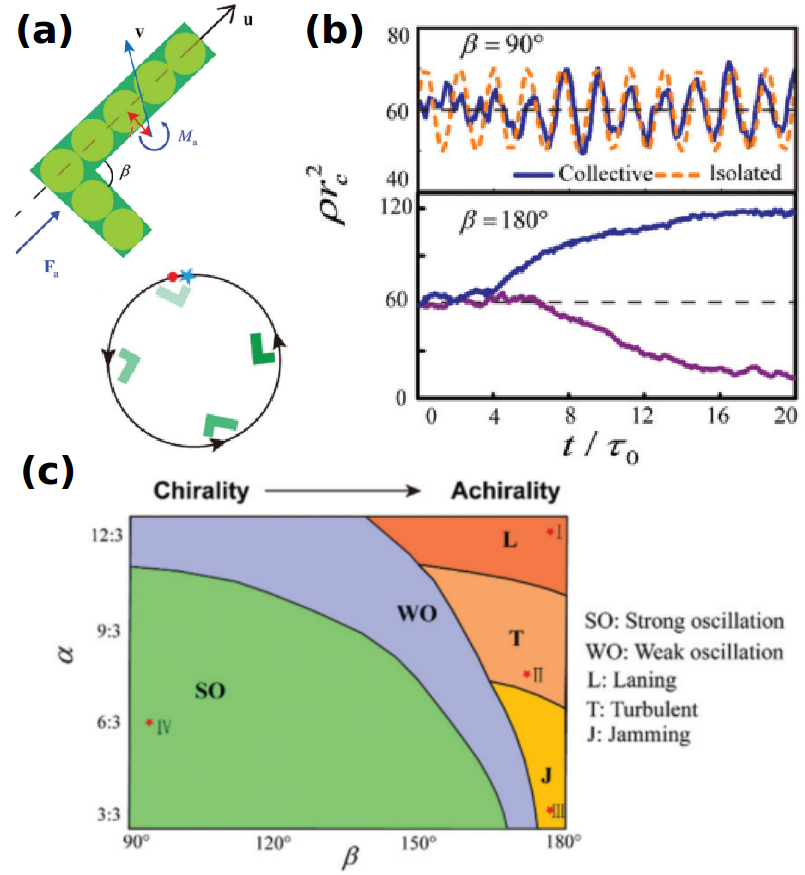}
\caption{\label{FigSPRLshapeDiagram} (a) Sketch of chiral L-shaped particle model and its circular motion. (b) Local number density as a function of time in the chiral rod system (top) and the achiral rod system (bottom). (c) Phase diagram with varying $\alpha$ and $\beta$ at packing fraction $\phi=0.2$~\cite{liu2019collective}.
Reproduced with permission.\cite{liu2019collective} Copyright 2019, Royal Society of Chemistry.
}
\end{figure}

Beyond the SPR model, if the shape of active particles is asymmetric, e.g. L-shaped, the eccentric self-propulsion induces an extra torque, which makes the active particle swim in a chiral circular fashion~\cite{lowenprecircle,lowenprlcircle}.
In Fig.~\ref{FigSPRLshapeDiagram}a, the L-shaped model is sketched, where $\alpha$ is the ratio between the length of long and short arms, and $\beta$ is the angle between the two arms. With an active force and torque, an isolated particle swims in a circular trajectory with the period $T_0$ increasing with $\alpha$ and $\beta$, so does the chirality.
In Ref.~\cite{liu2019collective}, an oscillation state was found with increasing the chirality, which is characterized by the periodic change of local density.

\subsection{Explicit alignment model}
\label{SECVicsekLike}
One of the major differences between SPRs and self-propelled spheres is that the steric effect leads to the effective alignment of propulsion direction via collision events between SPRs. Therefore, models of point like particles with explicit alignment interactions were applied to investigate the alignment effect, which are known as Vicsek-like models. Besides the steric hindrance, the aligning interactions may have other origins, like hydrodynamic interaction in systems of bacteria and actin filaments, and more complex for bird flocks or fish schools. In this section, we summarize the phase behaviour of these explicit alignment models.

One basic active particle model with alignment interaction is the Vicsek model (VM) first introduced to study the collective motion of self-propelled particles, in which particles are modeled as points without steric interaction~\cite{vicsek1995novel}. Based on the symmetry of alignment and self-propulsion, the self-propelled point models can be categorized into four kinds: polar point (PP/VM), polar rod (PR), apolar rod (AR, also called active nematics), Vicsek-shake (VS). General equations of motion for them are
\begin{eqnarray}
\begin{aligned}
	\mathbf{r}_j^{t+1} &=\mathbf{r}_j^t+v_{\pm}\mathbf{e}_{\theta_j}^{t'},\\
	\theta _j^{t+1} &=F_j(\theta ,\mathbf{r})+\eta \xi _j(t),
\end{aligned}
\end{eqnarray}
where $\mathbf{r}_{j}^t$ and $\theta_j^t$ are the position and orientation of particle $j$ at time $t$, and $\mathbf{e}_{\theta_j}^t$ is the unit vector pointing to the direction of $\theta_j^t$. Here $t'=t$ is backward difference and $t'=t+1$ is forward difference, and for VM there is no significant differences between the two methods~\cite{chate2008collective}.
$\xi_j(t)$ is a Gaussian white noise with zero mean and unit variance, and $\eta$ is the strength of noise.
PP and PR are polar driven, and $v_{\pm}=v_0$ is a constant propulsion velocity, while AR and VS are apolar driven, and $v_{\pm}=\pm v_0$ is chosen with equal probability; the function $F_j(\theta, \mathbf{r})$ indicates polar (PP, VS) and nematic (PR, AR) alignment interaction:
\begin{eqnarray}
\begin{aligned}
	\text{PP, VS:}~F_j(\theta,\mathbf{r}) &=\text{arg} \left[ \displaystyle{\sum_{k\in S_j}} e^{i\theta _k(t)} \right],\\
	\text{PR, AR:}~F_j(\theta,\mathbf{r}) &=\text{arg} \left[ \displaystyle{\sum_{k\in S_j}} \text{sign}[\cos (\theta_k(t)-\theta_j(t))] e^{i\theta _k(t)} \right],
\end{aligned}
\end{eqnarray}
where $S_j$ contains all the neighbours of particle $j$.
As a result of the balance between alignment and noise, decreasing $\eta$ leads to a disorder-order transition in all these four models
indicated by the increase of mean velocity. Because of different symmetries, these models exhibit various ordered states and phase transitions, which have been widely employed to study the collective motion of bird flocks, fish schools, bacteria, vibrated granular rods, microtubule-molecular motor suspensions, etc.

\begin{table*}[!ht]
\centering
\caption{Emergent phases in Vicsek-like models. }
\label{TablePhaseVM}
\begin{tabular}{c|c|c|c|c}
	\hline \multirow{2}{*}{Model} & \multirow{2}{*}{States (Decrease noise)} & \multicolumn{2}{c|}{Homogeneous ordered phase} & \multirow{2}{*}{Transition} \\ \cline{3-4}
	 & & Orientational Order & GNF exponent & \\
	\hline  {Polar Points~\cite{vicsek1995novel, gregoire2004onset, chate2008collective,solon2015phase}} & \makecell{I: disordered homogeneous\\ II: travelling bands (L-G coexistence)\\ III: ordered homogeneous} &  {Long range polar} & \multirow{2}{*}{$\sim 1.6$} & \makecell{I-II:  {1st order transition}\\ II-III:  {1st order transition}}\\
	\cline{1-3} \cline{5-5}  {Polar Rods~\cite{ginelli2010large}} & \makecell{I: disordered homogeneous\\ II: disordered unstable band\\ III: ordered bands\\ IV: ordered homogeneous} &  {Long range nematic} & & I-II:  {1st order transition}\\
	\hline  {Vicsek-shake~\cite{mahault2018self}} & \makecell{I: disordered homogeneous\\ II: ordered homogeneous} &  {Quasi-long range polar} & $\sim 1.73(3)$ &  {2nd order transition}\\
	\hline  {Apolar Rods~\cite{chate2006simple, ngo2014large}} & \makecell{I: disordered homogeneous\\ II: disordered unstable bands\\ III: ordered homogeneous} &  {Quasi-long range nematic} & $\sim 2.0$ & \makecell{ {Berezinskii–Kosterlitz–Thouless}\\  {transition}}\\
	\hline
\end{tabular}
\end{table*}

Table.~\ref{TablePhaseVM} shows the states observed in the four different models. For the PP model, decreasing the noise strength $\eta$ induces a phase transition from disordered homogeneous state to an ordered traveling bands state. Further decreasing $\eta$ leads to a long-range (LR, or global) polar ordered homogeneous phase with GNF, and the finite-size analysis suggests the disorder-order transition is discontinuous~\cite{vicsek1995novel, gregoire2004onset, chate2008collective}. Instead of focusing on the disorder-order transition of the global orientational order,
Ref.~\cite{solon2015phase} suggested that it is better to understand this transition as a gas-liquid transition, while the band solution is in a coexistence region, supported by the band fraction being proportional to excess density, which is consistent with the lever rule.
For the PR model, decreasing $\eta$ leads the disordered homogeneous state first to an unstable band state without global orientational order, further to a band state with long range nematic order, and finally to a homogeneous nematic ordered state with GNF~\cite{ginelli2010large}. Similarly, for the AR model, decreasing $\eta$ induces a chaotic disordered state, finally to a homogeneous quasi-long-range (QLR) nematic order state with GNF \cite{chate2006simple, ngo2014large}. For the VS model~\cite{mahault2018self}, the instability of homogeneous disordered phase is absent, decreasing $\eta$ directly leads to a homogeneous ordered phase, which exhibits a QLR polar order.

Even through these Vicsek-like models consider basic symmetry factors and facilitate theoretical works, there are still significant difference between the
Vicsek-like models and the realistic active particle models, which exhibit more complex phase behaviours as in Sec.~\ref{SECstericSPR}.
To bridge between these two different types of models, point particles with density dependent motility~\cite{farrell2012pattern} and lattice models, where each node can be occupied by at most one particle~\cite{peruani2011traffic}, were introduced to mimic the steric hindrance.
In the models above, a number of new patterns were found, and they partially recover  the phases observed in steric models, which are absent in point models, e.g., the polar ordered~\cite{farrell2012pattern} and glider states~\cite{peruani2011traffic} resembling the swarming phase~\cite{wensink2012emergent}, and the aster~\cite{farrell2012pattern} and traffic jam states \cite{peruani2011traffic} resembling the aggregate state \cite{weitz2015self} to some extent.
Most significantly, in the Vicsek model, the particles move perpendicular to the band, but in Ref.~\cite{farrell2012pattern, peruani2011traffic}, the band state recovers the parallel orientation as in volume excluded models. Especially in Ref.~\cite{farrell2012pattern}, the similarity of phase boundary between the homogeneous ordered phase and the disordered cluster phase obtained by simulation and linear instability analysis in continuum theories suggests that the cluster instability is determined by the effective ``pressure term'' $-\frac{1}{2}\nabla(v(\rho)\rho)$. This is similar to the MIPS scenario and mostly steric effect ~\cite{tailleur2008statistical, fily2012athermal, redner2013structure, farrell2012pattern}. In contrast, in the point model, adding a repulsive interaction by introducing an anti-alignment rotation to nearby particles inside a repulsion zone instead of the explicit excluded volume effect only leads to the conventional order-disorder transition~\cite{romenskyy2013statistical}.
A simple explanation for the differences observed in explicit alignment and steric models is that, in the latter, the traffic jam can trap particles and lead to denser disordered aggregates, while in the former, only the ordered region can sweep more and more particles leading to denser groups.
Besides adding effective repulsive interaction to Vicsek-like model, to bridge the gap from another side, efforts were also made to understand the effective alignment resulted from steric effects. In Ref~\cite{shi2018self}, by introducing more control parameters to the simulation of SPRs like the softness of repulsive potential and the anisotropy of motility, a phase diagram was constructed to encompass both collective dynamics corresponding to Vicsek-like model and MIPS. And recently, spontaneous local velocity alignment was also found in the dense phase of MIPS ~\cite{caprini2020spontaneous}. By using a suitable change of variables,  an effective Vicsek-like interaction was derived, which qualitatively predicts the  dependence of correlation lengths on activity.
The analytical description of velocity dynamics may also help understand the emergent polar order in anisotropic active colloids, where the alignment originates from steric effects.

\subsection{Order parameter and phase characterization}
\label{SECCharacter}
\subsubsection{Density field}
\begin{figure}[!ht]
\centering
\includegraphics[width=1.\linewidth]{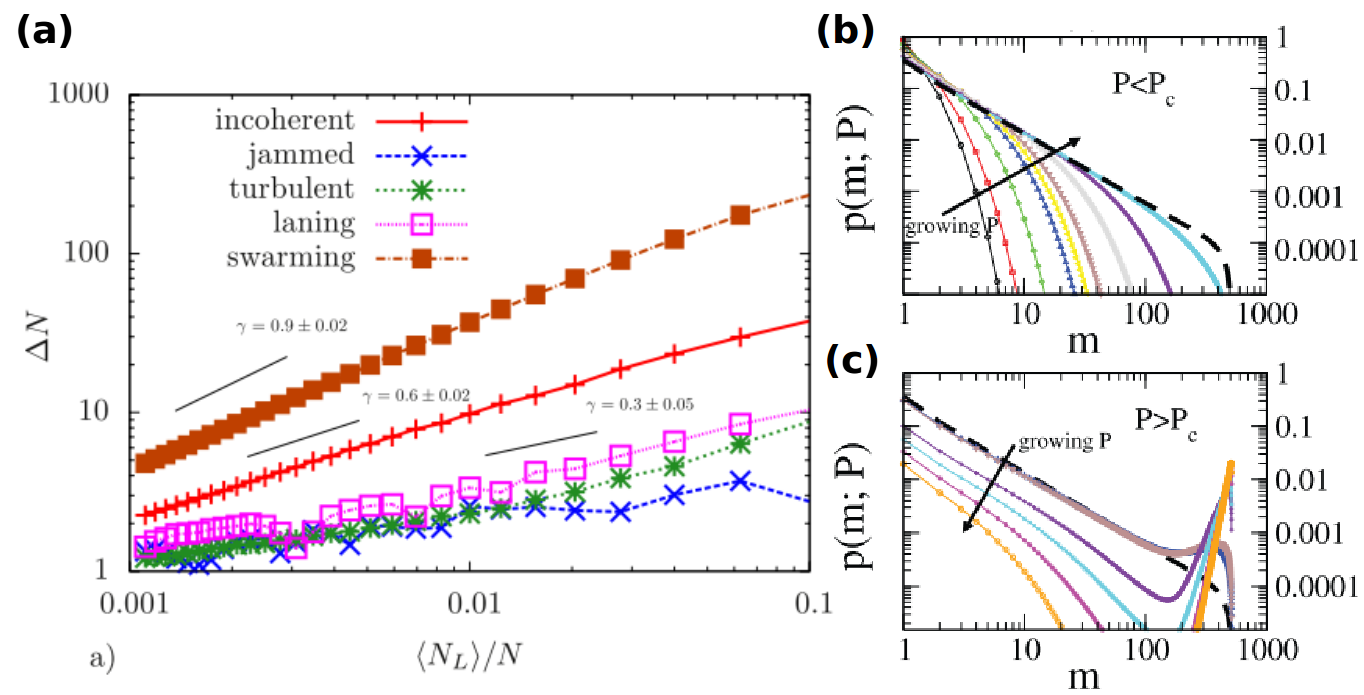}
\caption{\label{FigNumFluctStericSPR} (a) Number fluctuation of different phases for 2D steric SPR model \cite{wensink2012emergent}, and here $\gamma = \alpha /2$.
Reproduced with permission.\cite{wensink2012emergent} Copyright 2012, IOP Publishing.
 (b, c) Two kinds of cluster size distribution at two sides of critical weight of coagulation vs fragmentation $P_c$ ~\cite{peruani2013kinetic}. 
Reproduced with permission.\cite{peruani2013kinetic} Copyright 2013, IOP Publishing.
 }
\end{figure}

Different from the MIPS observed in spherical ABP systems, the density field in SPR systems exhibits a signature of GNF, which was first predicted by continuum theories~\cite{ramaswamy2003active}.
The number fluctuation of particles in a finite box of length $l$ is defined as $\sigma_l^2=\langle N_l^2 \rangle-\langle N_l \rangle ^2$.
According to the central limit theorem, $\sigma_l^2 \sim \langle N_l \rangle ^{\alpha}$ with the exponent $\alpha=1$ for randomly distributed particles. In contrast, $\alpha>1$ is called GNF indicating the existence of long-range correlation in the system.
Other properties that can be used to characterize the density field include the coarse-grained density probability distribution $P(\rho_l)$, the spatial density-density correlation function $C(\mathbf{r},t)=\langle \rho(\mathbf{0},t) \rho(\mathbf{r},t) \rangle$, and the structure factor $S(\mathbf{k})=\frac{1}{N}\sum_i | \exp (-i\mathbf{k \cdot} \mathbf{r}_i ) |^2$.
In the steric SPR model~\cite{wensink2012emergent}, the incoherent phase at low $\phi$ has $\alpha \simeq 1.2$ a little larger than the criterion of GNF, and the swarming phase at low $\phi$ and large $a$ exhibits significant GNF with $\alpha \simeq 1.8$, while other high density phases, i.e., jammed, turbulent and laning phases, have $\alpha \simeq 0.6$, as shown in Fig.~\ref{FigNumFluctStericSPR}.

Ref.~\cite{dey2012spatial} discussed the relationship between GNF exponent, $C(r)$ and $S(k)$ for different Vicsek-like models. In general, the number fluctuation information is embedded in the density correlation function $C(r)$, and
\begin{eqnarray}
\label{EqCrNumFluct}
	\sigma_{l}^{2}(t)=l^{d} \int_{0}^{l} \mathrm{d}^d \mathbf{r}\left[C(\mathbf{r}, t)-\langle\rho\rangle^{2}\right].
\end{eqnarray}
The results are summarized in Table.~\ref{TableDensityField}.
For PP and PR models, which are both polar driven, $S(k,t)\mathcal{L}^2(t)$ consists of two distinct power-law scalings for small and large $k\mathcal{L}(t)$, where $\mathcal{L}(t)$ is a macroscopic length scale over which enhanced clustering exists.
The small $k$ part corresponds to the long range density correlation tail $C(\mathbf{r}) \sim |\mathbf{r}|^{-\eta}$ for $r/\mathcal{L}\geq 1$, where $\eta=0.8$. According to Eq.~\ref{EqCrNumFluct}, the GNF is dominated by the integration near upper bound and has $\alpha=2-\eta/d=1.6$ with $d$ the dimensionality of the system. While for the apolar-driven rod (AR) model, there is no long power law tail in $C(r)$, so the integration is dominated by the cusp $C(r)\sim a_1-b_1|r/\mathcal{L}|^{\beta_1}$ in the range $r\ll \mathcal{L}$, which leads to the number fluctuation following $\sigma_{l}^{2} \sim a_1\langle N\rangle^{2}-\frac{b_1}{\mathcal{L}^{\beta_{1}}}\langle N\rangle^{\beta_{1} / d+2}+\cdots$, where $a_1$, $b_1$ and $\beta_1$ are positive constants.
\begin{table*}[!ht]
\centering
\caption{Relation between $C(r)$, $S(k)$ and $\sigma_l^2$ in active matter systems~\cite{dey2012spatial}.
Reproduced with permission.\cite{dey2012spatial} Copyright 2012, American Physical Society.
} \label{TableDensityField}
\begin{tabular}{p{1cm} | p{1.2cm} | p{1.2cm} | c | c | c}
	\hline Model & Align & Driven & $S(k)$ & $C(r)$ & $\sigma_l^2$ \\ [1ex]
	\hline Polar Point & Polar & Polar & \parbox[c][3.5cm]{4.5cm}{\includegraphics[width=4.2cm]{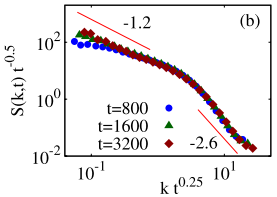}} & \multirow{2}{*}{\makecell{$C(\mathbf{r}) \sim |\mathbf{r}|^{-\eta}$ \\ \\ $\frac{r}{\mathcal{L}} \geq 1$}} & \multirow{2}{*}{\makecell{\includegraphics[width=4.5cm]{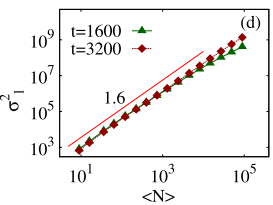}\\ \\ $\alpha =2-\frac{\eta}{d}$}} \\
	\cline{1-4} Polar Rod & Nematic & Polar & \parbox[c][3.5cm]{4.5cm}{\includegraphics[width=4.2cm]{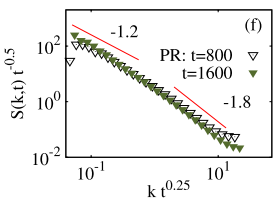}} & & \\
	\hline Apolar Rod & Nematic & Reversal & \parbox[c][4.5cm]{4.5cm}{\includegraphics[width=4.2cm]{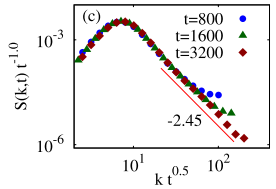}} & \makecell{\includegraphics[width=4cm]{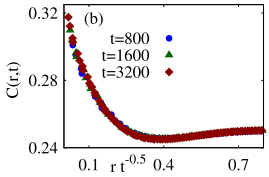} \\ \\ $C(r)\sim a_1-b_1\left| \frac{r}{\mathcal{L}} \right| ^{\beta_1}$ \\ $r \ll \mathcal{L}$} & \makecell{\includegraphics[width=4.5cm]{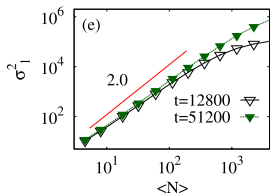} \\ \\ $\sigma_l^2 \sim a_1\langle N \rangle^2-\frac{b_1}{\mathcal{L}^{\beta_1}}\langle N \rangle ^{\beta_1/d+2}+...$} \\
	\hline
\end{tabular}
\end{table*}

Cluster size distribution (CSD) is another property used to study the density field. Ref.~\cite{peruani2010cluster, peruani2013kinetic} constructed a kinetic clustering equation for SPRs based on the control parameter $P$, which is the relative weight of coagulation with respect to fragmentation. It was found that there is a critical $P_c$ where the power law CSD $p(m;P_c)\sim m^{-\gamma}$, with $m$ the cluster size, separates an individual and a collective (clustering) phases.
For $P<P_c$, CSD can be well fitted with $p(m,P)\sim m^{-\gamma} \exp (-m/\tilde{m}(P))$ with $\tilde{m}(P)$ the characteristic cluster size at $P$, while for $P>P_c$, CSD is non-monotonic and has a bump for relatively large cluster size, as shown in Fig.~\ref{FigNumFluctStericSPR}.
Similar CSD behaviors below, at and above transition threshold are observed in various experimental \cite{peruani2012collective} and simulation \cite{peruani2006nonequilibrium, chate2008collective, yang2010swarm, Romenskyy2013} works of spherical ABP and SPR systems. Besides, the phase with a fat tails in the CSD appears always leading to GNF, while GNF does not always produce a fat tail. 
Thus a solid connection between GNF and the fat tails in the CSD remains to be established~\cite{peruani2013kinetic}.

\subsubsection{Orientational field}
The order parameter of orientational field is defined as $S_n=\langle | \langle \exp (in\theta_j) \rangle | \rangle_t$, and the orientational spatial correlation function is defined as $g_{n}(r)=\langle \cos \{ n[\theta(0)-\theta(r)] \} \rangle$, where $n$ is a positive integer, and $n=1$ and 2 for polar and nematic orders, repsectively.
The increase of $S_n$ implies the change from a disordered state to an ordered state.
And whether the ordered state has long-range order (LRO) or quasi-long-range-order (QLRO) can be studied from the finite size scaling of $\langle S_n \rangle$, the decay of $g_n(r)$ and whether the probability distribution of order parameter $P(S_n)$ converge to the Bramwell-Holdsworth-Pinto (BHP) distribution \cite{ngo2014large, bramwell2000universal, bramwell2001magnetic}.

\begin{figure}[!ht]
\centering
\includegraphics[width=1.\linewidth]{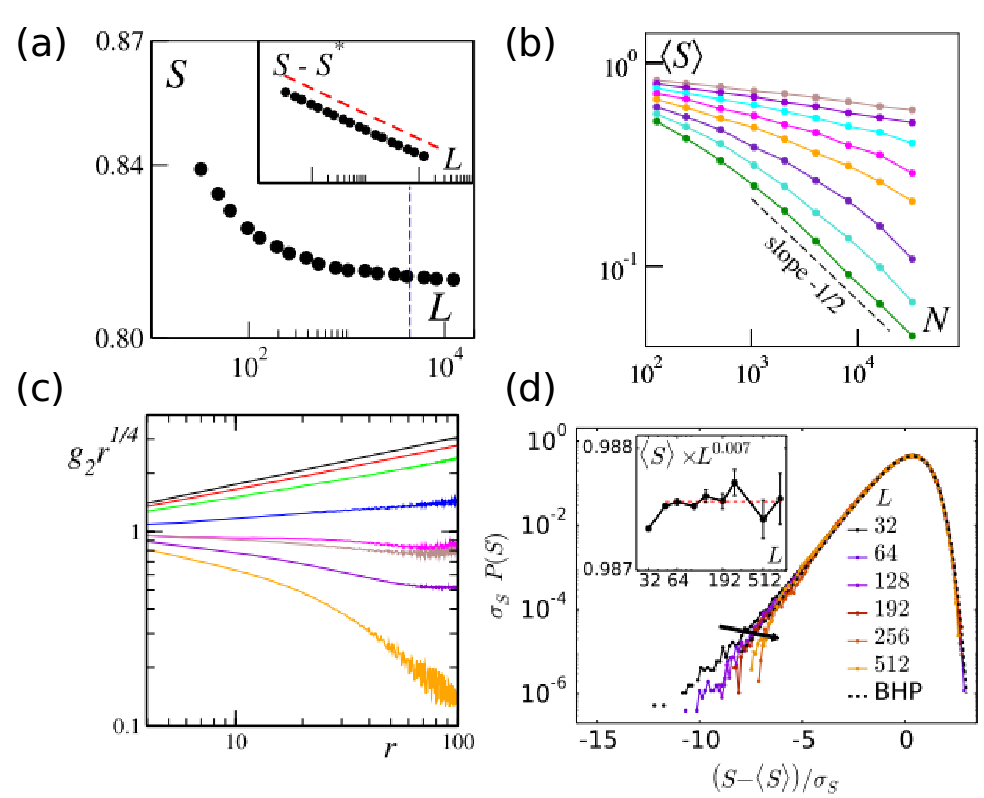}
\caption{\label{FigOrderFSARPR} (a) Long-range nematic order phase of PR model: nematic order parameter decays with system size slower than power law and approaches an asymptotic value $S^{*}$. \cite{ginelli2010large}
Reproduced with permission.\cite{ginelli2010large} Copyright 2010, American Physical Society.
 (b-d) Quasi-long-range nematic order phase of AR model: (b) the nematic order parameter decreases algebraically with system size $\langle S \rangle \sim N^{-\zeta (\eta)}$, with $\zeta$ increasing continuously with $\eta$, and crossover to fully disordered phase characterized by $\zeta=1/2$ at large-enough sizes \cite{chate2006simple}; (c) orientational correlation function $g_2(r) \sim r^{-1/4}$ at critical $\eta_c$ \cite{chate2006simple}; 
Reproduced with permission.\cite{chate2006simple} Copyright 2006, American Physical Society.
(d) rescaled distribution $P(S)$ at different system size converge to BHP distribution. Inset: $\langle S \rangle \times L^{-\zeta}$ vs $L$ \cite{ngo2014large}. 
Reproduced with permission.\cite{ngo2014large} Copyright 2014, American Physical Society.
}
\end{figure}

For ordered phases in PP and PR models, the finite size scaling of $S_n$ decays slower than a power law and approaches a finite asymptotic value $S^*$ as shown in Fig.~\ref{FigOrderFSARPR}a, suggesting a LRO. For the VS and AR models, the finite size scaling of $S_n$ shows a crossover of effective exponent $\zeta(\eta,N)=-\frac{\dd \ln S}{\dd \ln N}$ from a small finite value towards $1/2$, suggesting a fully disordered state at $N\rightarrow \infty$ (Fig.~\ref{FigOrderFSARPR}b), which resembles the QLRO in an equilibrium XY model. Besides, the $g_2(r) \sim r^{-1/4}$ at the threshold $\eta_c$ in VS and AR models is also similar with the scaling of XY model. Furthermore, as $L$ increases, the order parameter  distribution $P(S_n)$ converges to BHP distribution, as shown in Fig.~\ref{FigOrderFSARPR}d, which also suggests a QLRO~\cite{ngo2014large}.

\begin{figure}[!ht]
\centering
\includegraphics[width=1.\linewidth]{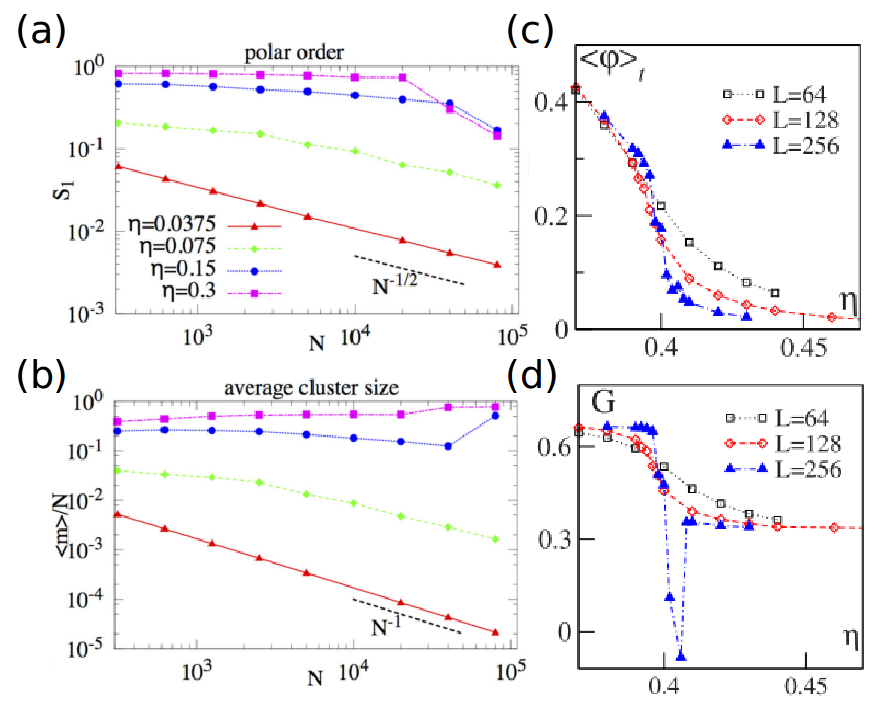}
\caption{\label{FigFSOrientOrder} (a-b) Finite size scaling of polar order parameter and average cluster size in self-propelled rods with steric interaction, where global polar order perishes at relative large $N$ while aggregate size keep increasing linearly with $N$ \cite{weitz2015self}. 
Reproduced with permission.\cite{weitz2015self} Copyright 2015, American Physical Society.
(c-d) Order-disorder phase transition in PP model shows continuous and discontinuous manners at small and large $L$, respectively~\cite{chate2008collective}. 
Reproduced with permission.\cite{chate2008collective} Copyright 2008, American Physical Society.
}
\end{figure}

Both in the steric hindrance models and explicit alignment models, finite size analysis of the orientational order is necessary for determining the type of ordered states and disorder-order phase transitions.
For example, in Ref.~\cite{weitz2015self}, a global polar order is observed for relatively smaller system size $N$. However, when $N$ reaches $\sim 10^5$, $S_1$ decreases with $N$ (Fig.~\ref{FigFSOrientOrder}a) while the average cluster size keeps increasing linearly with $N$ (Fig.~\ref{FigFSOrientOrder}b).
In the PP model, at small $L$, the order-disorder phase transition appears to be continuous, which can be seen in the continuous change of $\langle \varphi \rangle$ (same as $S_1$) and the crossover of Binder cumulant $G(\eta)=1-\frac{\langle \varphi ^4 \rangle}{3\langle \varphi ^2 \rangle ^2}$ at a critical noise strength $\eta_c$ for different $L$. However, with increasing $L$, $\langle \varphi \rangle$ change sharply at certain threshold (Fig.~\ref{FigFSOrientOrder}c) and $G$ drops towards a negative value (Fig.~\ref{FigFSOrientOrder}d), which both indicate the discontinuous feature of the transition~\cite{chate2008collective}.

\subsubsection{Vortical states and turbulence}
In experiments, collective swirling motion of active units and vortical patterns have been observed in various systems, e.g., mixtures of actin or microtubules and motor proteins \cite{sanchez2012spontaneous}, swimming bacteria \cite{dombrowski2004self, wensink2012meso, wu2017transition}, vibrated rods \cite{blair2003vortices}, etc.  In computer simulations of SPRs, the turbulent state was qualitatively recovered \cite{saintillan2007orientational, wensink2012emergent}. In biological systems, the turbulent state optimizes fluid mixing, and the corresponding optimal aspect ratio of rods found in SPR simulations fall in the range of typical bacterial cells \cite{wensink2012meso}.
The turbulent state is characterized by using properties like velocity field structure, enstrophy density, and energy spectrum.

\begin{figure}[!ht]
\centering
\includegraphics[width=1.\linewidth]{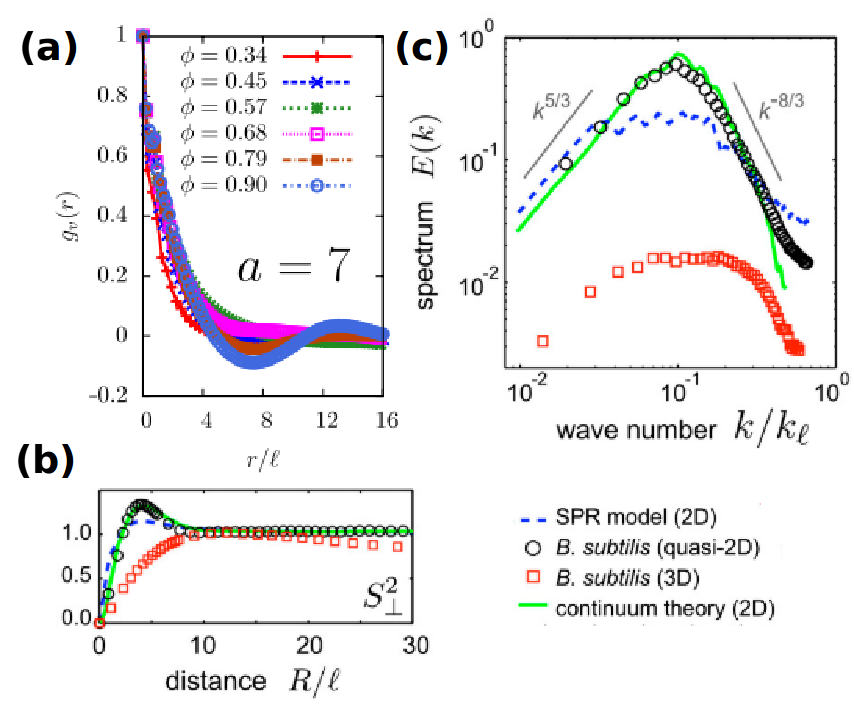}
\caption{\label{FigTurbulentChara} (a) Velocity correlation of SPR \cite{wensink2012emergent}. 
Reproduced with permission.\cite{wensink2012emergent} Copyright 2012, IOP Publishing.
(b) Second moment of transverse velocity correlation \cite{wensink2012meso}. (c) Energy spectrum of turbulent states in experiment(2D, 3D), SPR model and continuum model \cite{wensink2012meso}.
Reproduced with permission.\cite{wensink2012meso} Copyright 2012, National Academy of Sciences.
}
\end{figure}

\textbf{Velocity structure} Velocity correlation (VC) is defined as $g_v(r)=\langle (\mathbf{v}(0) \cdot \mathbf{v}(\mathbf{r})-\langle v \rangle^2 )\rangle /(\langle v^2 \rangle -\langle v\rangle ^2)$, where the minima of VC reflects a typical vortex size $R_v$, and the non-monotonic decrease of $g_v(r)$ with negative correlations represents more pronounced vortical motion (Fig.~\ref{FigTurbulentChara}a). More detailed structure can be revealed by different moments of transverse velocity structure functions $S^{n}_{\perp} (\mathbf{R})=\langle (\mathbf{v}_{\perp}(\mathbf{R},t)-\mathbf{v}_{\perp}(\mathbf{0},t))^n \rangle$ with $n$ the order of moments, where the maxima of the even transverse structure $S_{\perp}^{2n}$ also signals $R_v$ (Fig.~\ref{FigTurbulentChara}b).
In agent based simulations, it is found that $g_v(r)$ is insensitive to the bulk density and aspect ratio of particles~\cite{wensink2012emergent}, which suggests the possible existence of a universal vortical structure.

\textbf{Vorticity} In 2D systems, the enstrophy per unit area is defined as $\Omega=\frac{1}{2}\langle |w(\mathbf{r},t)|^2 \rangle$, which measures the kinetic energy associated with the local vortical motion. For slender rods, the mean enstrophy exhibits a pronounced maximum in the turbulent state. Besides, in hydrodynamic models, the probability distribution of velocity follows a Gaussian distribution, while for vorticity, as it is the function of the velocity gradient, its probability distribution is different from the Gaussian distribution due to the spatial correlation \cite{giomi2015geometry}.

\textbf{Energy spectrum} To compare with classical 2D turbulence, the energy spectrum is obtained through the Fourier transform of velocity correlation: $E(k)=\frac{k}{2\pi}\int \dd \mathbf{r} \exp^{-i\mathbf{k \cdot r}} \langle \mathbf{v}(0,t) \cdot \mathbf{v}(\mathbf{r},t) \rangle$, which is supposed to follow a power-law decay with $k^{-5/3}$  {at large $k$} predicted by the Kolmogorov-Kraichanan scaling theory \cite{kraichnan1967inertial, kraichnan1980two}. In contrast, for SPRs, energy is injected from microscopic length scale and a different $E(k)$ behavior is expected. In both experiment and continuum model, an exponent of $-8/3$ is observed \cite{wensink2012meso}, but in agent-based SPR model simulation an approximate $-5/3$ exponent is observed \cite{wensink2012emergent} ( {the blue dashed line in} Fig.~\ref{FigTurbulentChara}c). Besides, the 2D SPR model and 3D experimental systems show the intermediate plateau region indicating that kinetic energy is more evenly distributed over a range of length scales.

\section{Hydrodynamic effects on the dynamic assembly of active colloids}\label{HESEC}
Active colloidal particles, such as bacteria and artificial microswimmers, self-propel in an embedding fluid. Therefore, besides steric interparticle interactions, the fluid-mediated hydrodynamic interactions (HI) play an important role in the collective assembly of intrinsically nonequilibrium active colloids~\cite{lauga2009hydrodynamics}. Hydrodynamic interactions, which essentially arise from momentum and mass conservation of active suspensions, are long-ranged and have many-body effects, and sensitively depend on the type and shape of the microswimmer and on the dimensionality and boundary of the system. Thus, the hydrodynamic effects need to be taken in account when theoretically studying realistic active colloids.

Generally, the solvent effect on the active colloids can been addressed in two different ways: continuum theories and direct particle simulations. In continuum theories, the hydrodynamic effects are included by coupling the equations describing active particles to the Navier-Stokes equation, in which an active stress on the fluid exerted by the microswimmers is added~\cite{marchetti2013hydrodynamics}. With this idea, Aditi Simha and Ramaswamy studied the effect of HI on the collective motion of swimming rods~\cite{simha2002hydrodynamic}. They found that the nematic alignment of the self-propelled rods at low Reynolds number is always unstable to long wavelength disturbances and the system demonstrates giant number density fluctuations. Similar models have been extensively developed to investigate the orientational order, stability and collective motion of active suspensions~\cite{wolgemuth2008collective,aranson2007model,saintillan2008instabilities,baskaran2009statistical}.
Recently, a continuum theory of self-propelled particles (without alignment) in a momentum-conserving solvent is developed to study phase separation and coarsening dynamics~\cite{tiribocchi2015active,singh2019hydrodynamically}.
These continuum theories usually consider far-field hydrodynamics and do not capture the details of the HI and swimmers, and they are limited to very large length scales.

Alternatively, direct particle simulations treat the hydrodynamics of active suspensions by explicitly modeling microswimmers with the Navier-Stokes equation of the solvent solved via different  numerical techniques~\cite{zottl2016emergent,elgeti2015physics}. Such methods can properly capture complex interparticle interactions and the details of the swimming mechanisms, particle geometry and hydrodynamics, which are critical to semidilute and concentrated suspensions. Direct particle simulations mimic experimental systems more realistically, however its large-scale implementation requires high computational cost. In the following, we briefly review active colloidal models frequently used in the direct particle simulations.

\subsection{Spherical squirmers}

\begin{figure}[t]
\includegraphics[width=1.\linewidth]{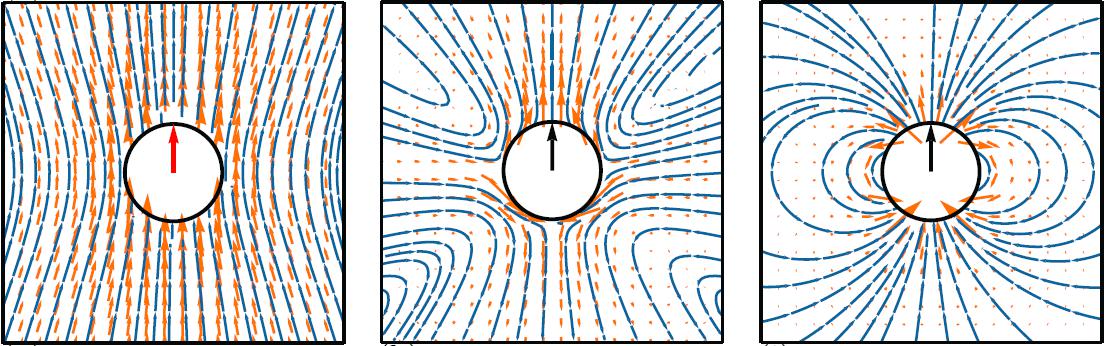}
\caption{Typical bulk velocity fields and streamlines around spherical particles: (Left) driven by an external force with far-field flow velocity decaying as $1/r$ with respect to the particle center, (Middle) a pusher squirmer of $\beta=-3$ with an $1/r^2$ decay, and (Right) a neutral squirmer with an $1/r^3$ decay~\cite{zottl2016emergent}. The flow field around a puller squirmer of $\beta=3$ is similar to the pusher but with the fluid pulled inwards along its body axis and pushed from its equator~\cite{zottl2016emergent}.
Reproduced with permission.\cite{zottl2016emergent} Copyright 2016, IOP Publishing.
}\label{Fig1}
\end{figure}

A simple model of swimmers called ``squirmers'' is introduced
by Lighthill~\cite{lighthill1952squirming}, refined further by
Blake~\cite{blake1971spherical}, and simulated for the first time by Ishikawa \emph{et al.}~\cite{ishikawa2007diffusion}. The squirmer was proposed to model ciliated microorganisms averagely. The simplest squirmer is a solid spherical particle of radius $R$ with a prescribed tangential surface velocity field~\cite{zottl2016emergent,ishikawa2006hydrodynamic},
\begin{equation}
{\bf v}_s({\bf \hat{r}_s})=B_1(1+\beta{\bf e}\cdot{\bf \hat{r}_s})({\bf e}\cdot{\bf \hat{r}_s}{\bf \hat{r}_s}-{\bf e}),\label{eq1}
\end{equation}
with ${\bf e}$ the particle orientation (swimming direction) and ${\bf\hat{r}_s}$ the unit vector of surface position from the particle center. Here, $B_1$ quantifies the self-propelling speed of an isolated
squirmer in the bulk, $v_{0}=2B_{1}/3$, and the coefficient $\beta$ captures the active stress with $\beta>0$, $\beta<0$ and $\beta=0$, respectively, corresponding to puller (pull fluid inwards along its body axis, e.g., Chlamydomonas), pusher (push fluid outwards along its body axis, e.g., E. coli) and neutral squirmer (e.g., Paramecium). This imposed surface velocity determines how the swimmer displaces in the embedding solvent with vanishing net force and torque. The representative flow fields around the squirmers are plotted in Fig.~\ref{Fig1}.

\begin{figure}[t]
\includegraphics[width=1.\linewidth]{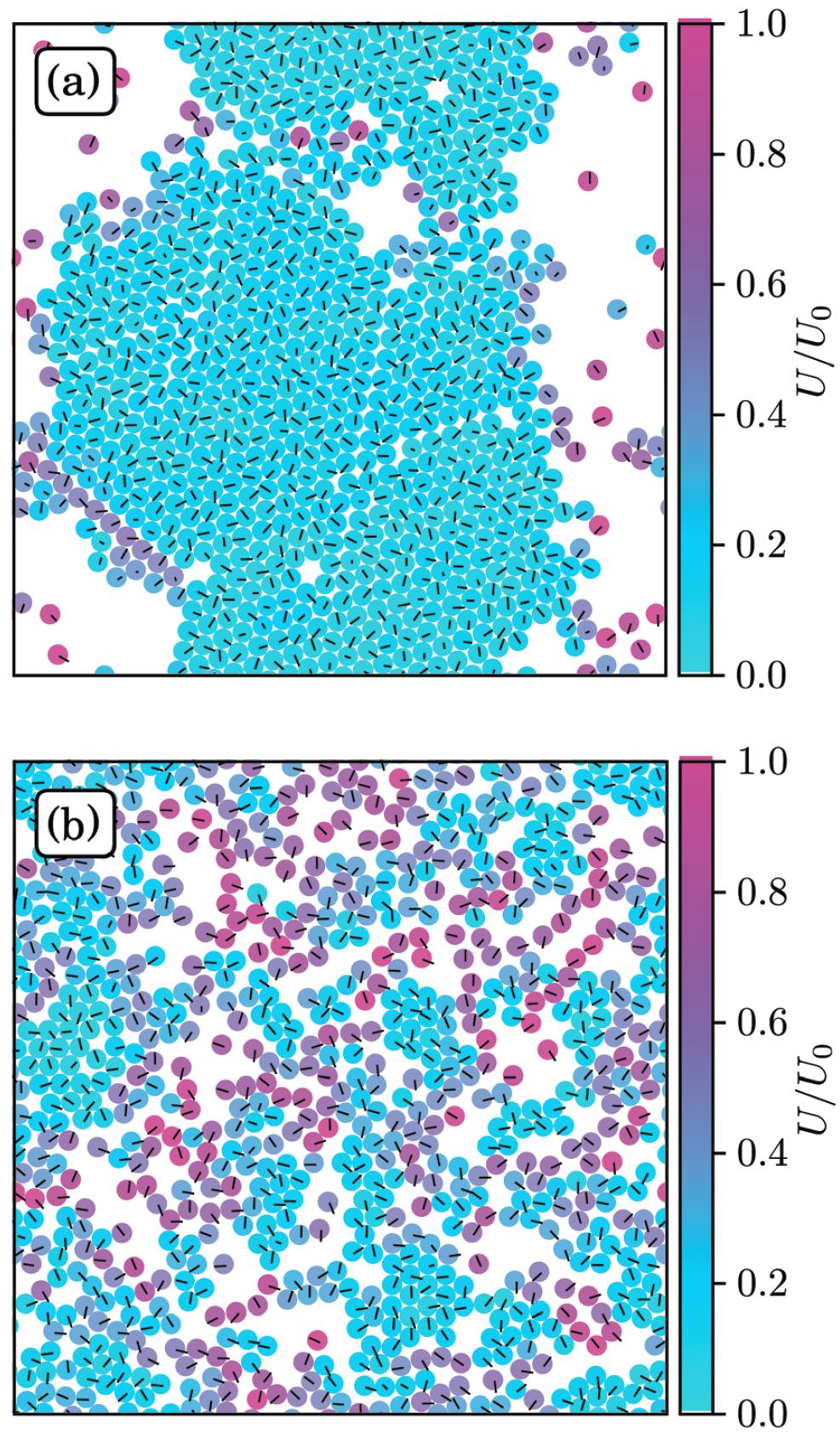}
\caption{Snapshots of spherical swimmers at the packing fraction $\phi=0.6$. (a) Active Brownian particles exhibit MIPS. (b) Neutral squirmers ($\beta=0$) only exhibit small dynamic clusters. The colors indicate instantaneous velocities of the swimmers~\cite{theers2018clustering}.
Reproduced with permission.\cite{theers2018clustering} Copyright 2018, Royal Society of Chemistry.
}\label{Fig2}
\end{figure}

The spherical squirmer system has been widely used to study hydrodynamic effects on the MIPS and motility-induced clustering (MIC), which has been better understood in dry active systems, as discussed in the above sections. The basic mechanism of the MIC and MIPS of sterically repulsive active particles is essentially a positive feedback between slowing-induced accumulation and accumulation-induced slowing. HI can change both the self-propelling velocity and the rotational dynamics of the microswimmer, hence are expected to influence the MIPS. In the absence of thermal fluctuations, numerical simulations of a 2D (repulsive) squirmer suspension~\cite{matas2014hydrodynamic} and a squirmer monolayer embedded in a 3D fluid~\cite{yoshinaga2017hydrodynamic} have shown that hydrodynamics strongly suppresses the MIPS across a wide range of values of $\beta$. This is because that HIs cause particles to undergo a large rotation each time they meet, so that the feedback argument fails. Particularly, the studies of Ref.~\cite{yoshinaga2017hydrodynamic} emphasized the importance of near-field HI in the suppression of the MIPS. Contrarily, a qusi-2D hydrodynamic simulation involving thermal fluctuations, where the squirmers are strongly confined by two walls but rotate freely in three dimensions, predicted HI enhanced clustering and phase separation of the spherical squirmers~\cite{zottl2014hydrodynamics,blaschke2016phase}. The phase behavior of squirmers depends on the swimmer type. In order to solve this contradiction, Theers \emph{et al.} simulated the same system ~\cite{zottl2014hydrodynamics,blaschke2016phase} but with much lower fluid compressibility~\cite{theers2018clustering}. They found no MIPS for the spherical squirmers in the quasi-2D system, as shown in Fig.~\ref{Fig2}, consistent with Ref.~\cite{matas2014hydrodynamic}. The discrepancy was attributed to peculiarities of the compressible fluid employed in Ref.~\cite{zottl2014hydrodynamics}. Hence, HIs suppress the MIPS and cluster formation of the spherical squirmers compared to ABPs. Moreover, the active stress parameter $\beta$ was found to correlate negatively with the suppression of cluster formation~\cite{theers2018clustering}.

However, for semidilute suspensions (volume fraction $\phi\simeq0.1$) of athermal squirmers in three dimensions, previous simulations~\cite{alarcon2013spontaneous,delmotte2015large,ishikawa2008development} have found a clear transient aggregation. Both semidilute athermal squrimer monolayer~\cite{ishikawa2008coherent} and 2D swimming disks~\cite{llopis2006dynamic} showed some clustering. These results indicate that HIs (particularly far-field hydrodynamics) is responsible for the MIC (instead of MIPS) for dilute situations in the absence of thermal fluctuations. The role played by thermal fluctuations need to be further clarified, and the effect of the dimensionality of the system also remains unsolved. Besides the significant influence on the structure of the squirmers, HI can often lead to a (local) polar order over a certain range of force-dipole strengths (especially for small $|\beta|$) in the spherical microswimmer suspensions~\cite{yoshinaga2017hydrodynamic,alarcon2013spontaneous,delmotte2015large,ishikawa2008coherent,ishikawa2008development,llopis2006dynamic,delfau2016collective,evans2011orientational}.
The developed polar order has a purely hydrodynamic origin, in contrast to the spherical ABP systems, in which the orientational order lacks. The combination of the clustering and the polar order may give rise to coherent motions of the squirmers and even oscillatory behaviors~\cite{alarcon2013spontaneous,oyama2016purely}.

\begin{figure}[b]
\includegraphics[width=1.\linewidth]{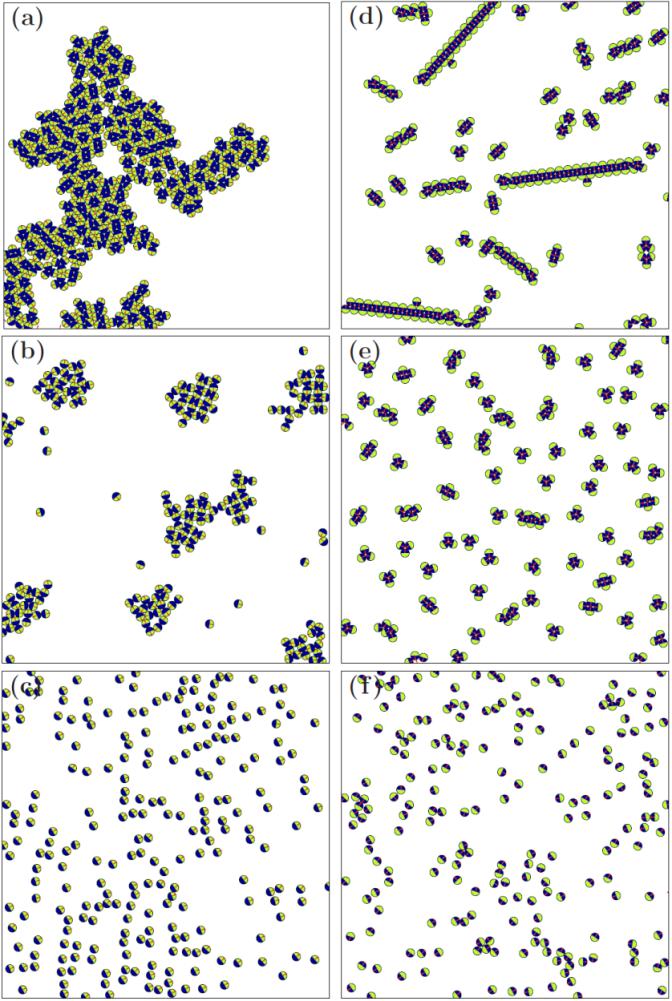}
\caption{Snapshots of different types of structures of the squirmers with anisotropic interactions. The attractive hemisphere is represented in blue, and the repulsive in green. (a) Coarsening, (b) Clustering, (c) Gas with polar order, (d) Chains, (e) Trimers state, (f) Gas with nematic order~\cite{alarcon2019orientational}.
Reproduced with permission.\cite{alarcon2019orientational} Copyright 2019, American Physical Society.
}\label{fig3}
\end{figure}

In addition, simulations of 2D athermal suspensions of isotropic attractive squirmers (with 2D or 3D hydrodynamics) showed an enhanced clustering~\cite{navarro2015clustering,alarcon2017morphology} due to the steric attraction. The properties of the clusters are sensitive to the competition between the self-propulsion, hydrodynamics and steric attraction. For low activities, the attractions dominate over the self-propulsion and the system exhibits an equilibrium phase behavior. Even in the presence of an attractive potential, HIs still suppress the phase separation. A monolayer of athermal squirmers with anisotropic (amphiphilic) interactions in a qusi-3D fluid showed quite rich structural behaviors~\cite{alarcon2019orientational} (Fig.~\ref{fig3}), in which the self-propulsion direction was tuned either towards the hydrophilic or hydrophobic side.

\begin{figure}[b]
\includegraphics[width=1.\linewidth]{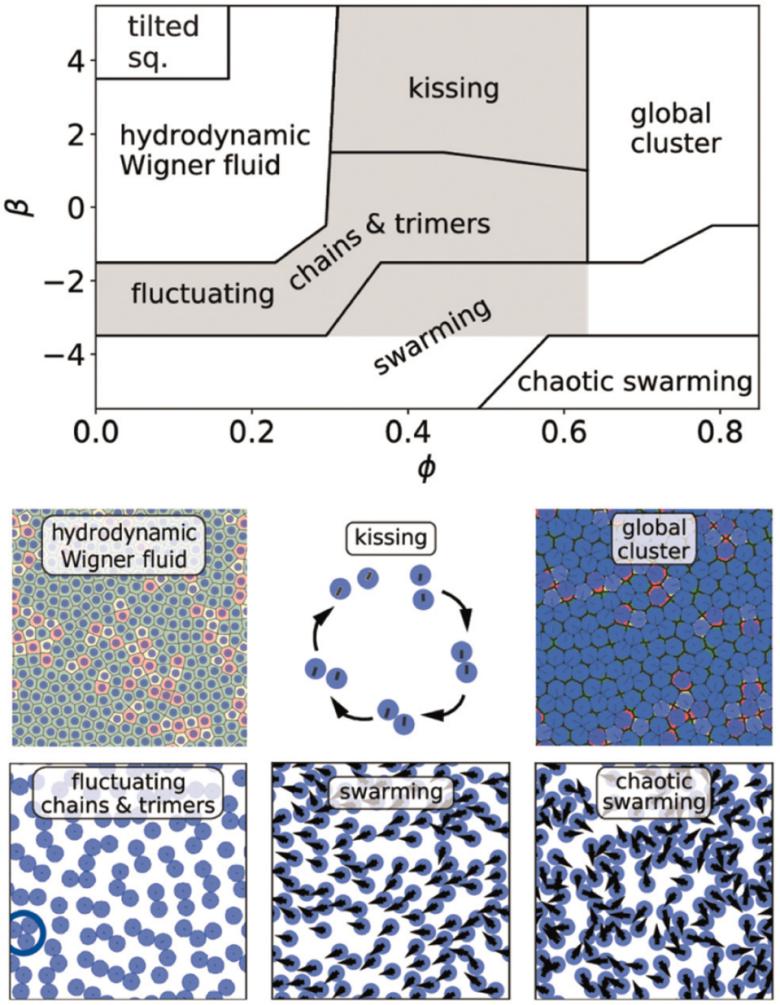}
\caption{Top: Schematic state diagram illustrating the collective dynamics
of a monolayer of squirmers confined to a bounding wall by gravity, with the squirmer type $\beta$ versus the area fraction $\phi$. Bottom: Snapshots of the observed dynamic states as seen from above. The hydrodynamic Wigner fluid (local hexagonal order with vanishing long-range translational and orientational order) is represented by a Voronoi tessellation. The kissing state is represented by a single kissing event. The ring in the fluctuating cluster state shows a trimer cluster. In the (chaotic) swarming state the arrows point along the orientation of the squirmers~\cite{kuhr2019collective}.
Reproduced with permission.\cite{kuhr2019collective} Copyright 2019, Royal Society of Chemistry.
}\label{fig4}
\end{figure}

In active colloidal suspensions, self-propelled particles of high density or bottom heaviness (nonuniform density distribution) can strongly accumulate to the substrate due to gravity or gravity-induced torque. If the stationary orientation of microswimmers is nearly normal to the boundary, the swimmer motility will be dramatically reduced. In this case, the swimmer drives a large fluid flow parallel to the boundary, which is able to entrain the surrounding particles. The swimmers interact with each other via their self-generated flow fields and may also phase separate~\cite{kuhr2019collective,rajesh2016universal,thutupalli2018flow, Lintuvuori1,Lintuvuori2}. A hydrodynamic simulation of a monolayer of squirmers confined to a boundary by gravity showed a variety of dynamic states (Fig.~\ref{fig4}) as the concentration and the squirmer type are varied~\cite{kuhr2019collective}. The swimmer aggregation driven by the self-generated flow is essentially different from the MIC, since the external force or torque is not free in this situation.

\subsection{Rod-like swimmers}

\begin{figure}[t]
\includegraphics[width=1.\linewidth]{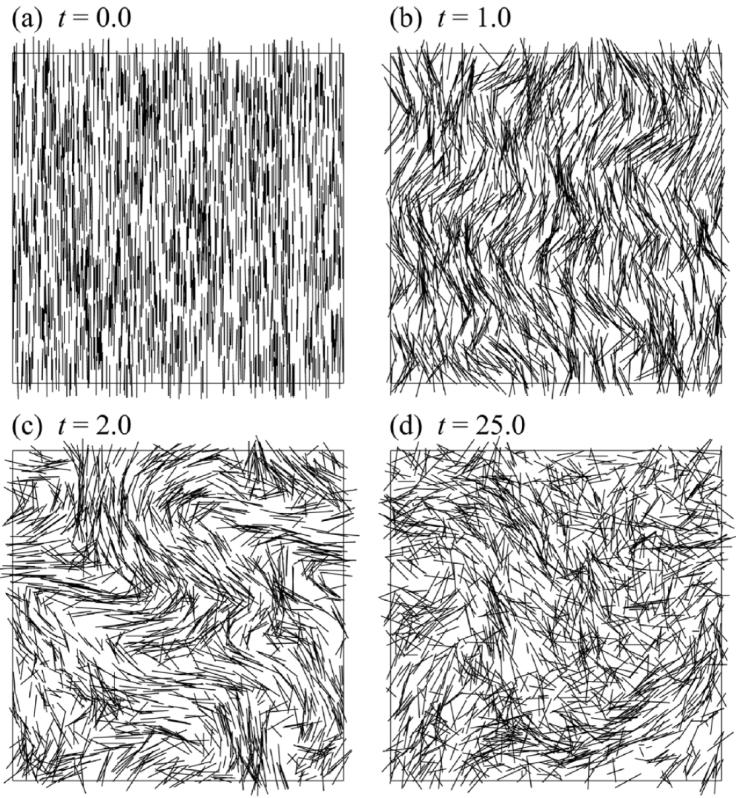}
\caption{Orientational instability and coherent motion in a suspension
of slender-rod pushers. The figure shows different stages of
the instability (a)-(d). The suspension quickly becomes isotropic from its initial aligned homogeneous state. Even at steady state, the microstructure of the suspension is clearly not random and demonstrates local orientation order~\cite{saintillan2007orientational}.
Reproduced with permission.\cite{saintillan2007orientational} Copyright 2007, American Physical Society.
}\label{fig5}
\end{figure}

Because of the anisotropic particle shape, rod-like microswimmers have different steric interactions and near-field hydrodynamics compared with spherical swimmers, and they experience intrinsic alignment couplings. Thus, rod-like swimmers exhibit different collective behaviors. A rod-like swimmer can be modelled as a rigid dumbbell with a phantom flagellum (It exerts a constant force along symmetric axis on one bead of the dumbbell and an equal and opposite force on the fluid.)~\cite{hernandez2005transport}, a slender rod with a prescribed surface shear stress~\cite{saintillan2007orientational}, or a spheroidal squirmer~\cite{theers2018clustering}.

\begin{figure}[t]
\includegraphics[width=1.\linewidth]{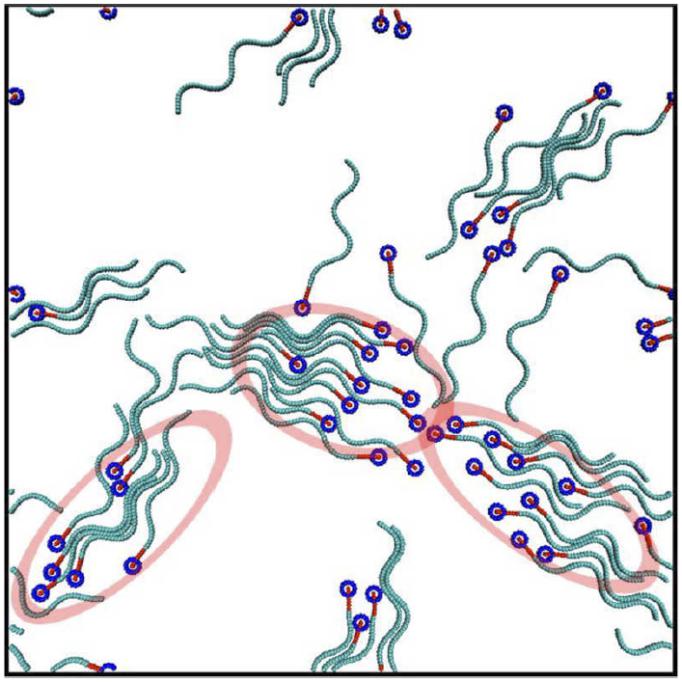}
\caption{Snapshot from simulations of 50 sperm in a 2D suspension with periodic boundary conditions. The red ellipses indicate large sperm clusters~\cite{yang2008cooperation}.
Reproduced with permission.\cite{yang2008cooperation} Copyright 2008, American Physical Society.
}\label{fig6}
\end{figure}

Unlike spherical squirmer systems, where puller or neutral squirmers are easier to form orientation order, pusher or neutral rod-like swimmers often show local orientation order and form large-scale dynamic structures. Neglecting thermal fluctuations and based on far-field approximation, Hernandez-Ortiz \emph{et al.}~\cite{hernandez2005transport} showed that HIs among 3D dumbbell pushers cause coherent fluid motions with a characteristic length much larger than the particle size. Similarly, Saintillan \emph{et al.}~\cite{saintillan2007orientational,saintillan2011emergence} reported that for slender-rod pushers an isotropic or aligned homogeneous state becomes unstable due to HIs, and a local orientation ordering develops and a large-scale correlated motion emerges (Fig.~\ref{fig5}), which was not observed in puller suspensions. Recent simulations of an ellipsoidal squirmer monolayer with 3D hydrodynamics involving the near-field effects showed similar collective motions~\cite{kyoya2015shape}.
Yang \emph{et al.} studied the hydrodynamic swarming of 2D sperms~\cite{yang2008cooperation} and swimming flagella~\cite{yang2010swarm}, and observed the formation of dynamic clusters with local polar order, as shown in Fig.~\ref{fig6}. When two sperms (rod-like pusher) with the same beat frequency happen to get close together and swim in parallel, due to HIs they synchronize, attract and swim together.

\begin{figure}[b]
\includegraphics[width=1.\linewidth]{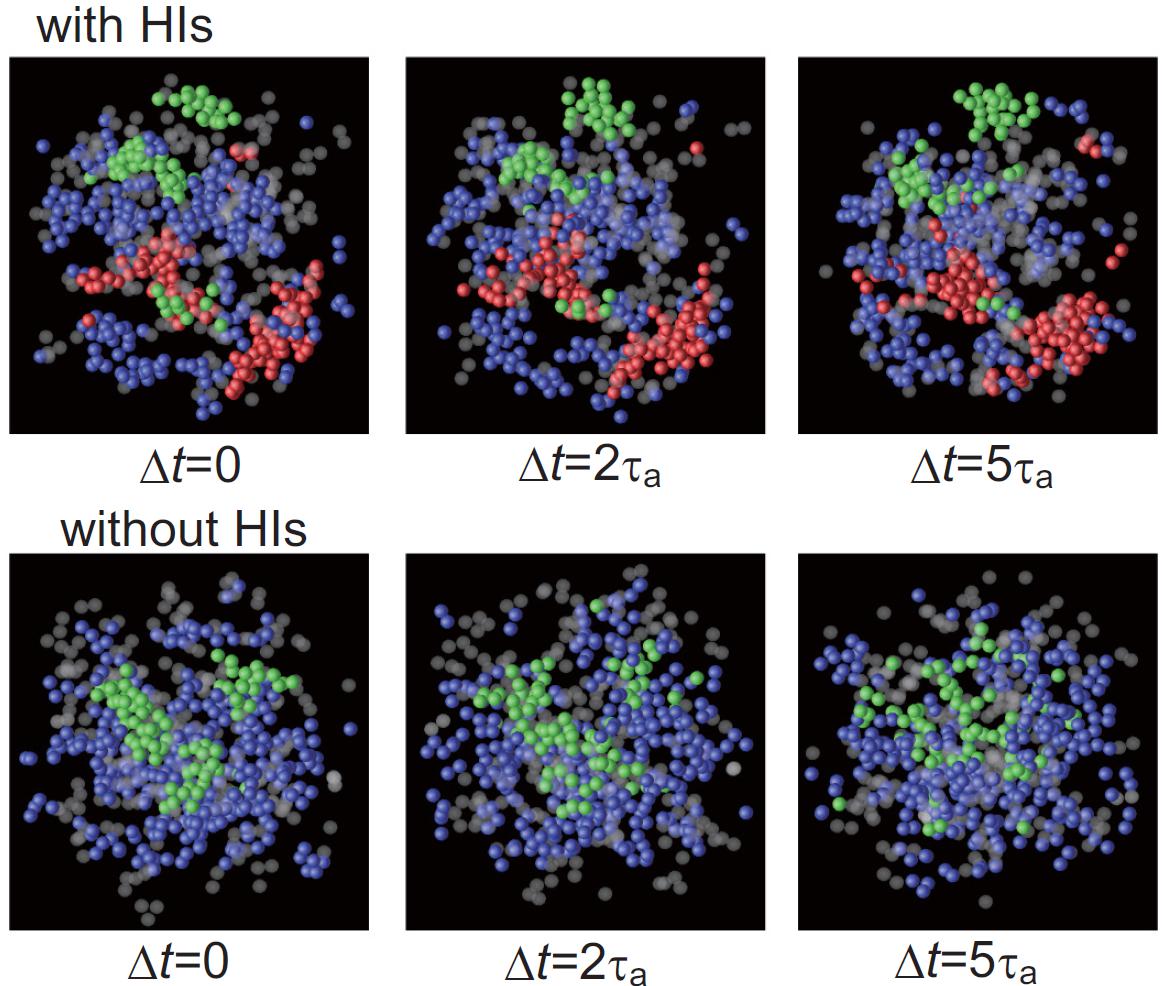}
\caption{The typical time evolution of the cluster configuration in the steady state with (Top) and without (Bottom) HI in semidilute dumbbell systems. Here, different colors represent different numbers of dumbbell swimmers in a cluster ($N_c$): blue for $2\leq N_c<10$, green for $10\leq N_c<20$, and red  $N_c\geq20$. The isolated swimmers are shown as grey semitransparent dumbbells~\cite{furukawa2014activity}.
Reproduced with permission.\cite{furukawa2014activity} Copyright 2014, American Physical Society.
}\label{fig7}
\end{figure}

\begin{figure}[t]
\includegraphics[width=1.\linewidth]{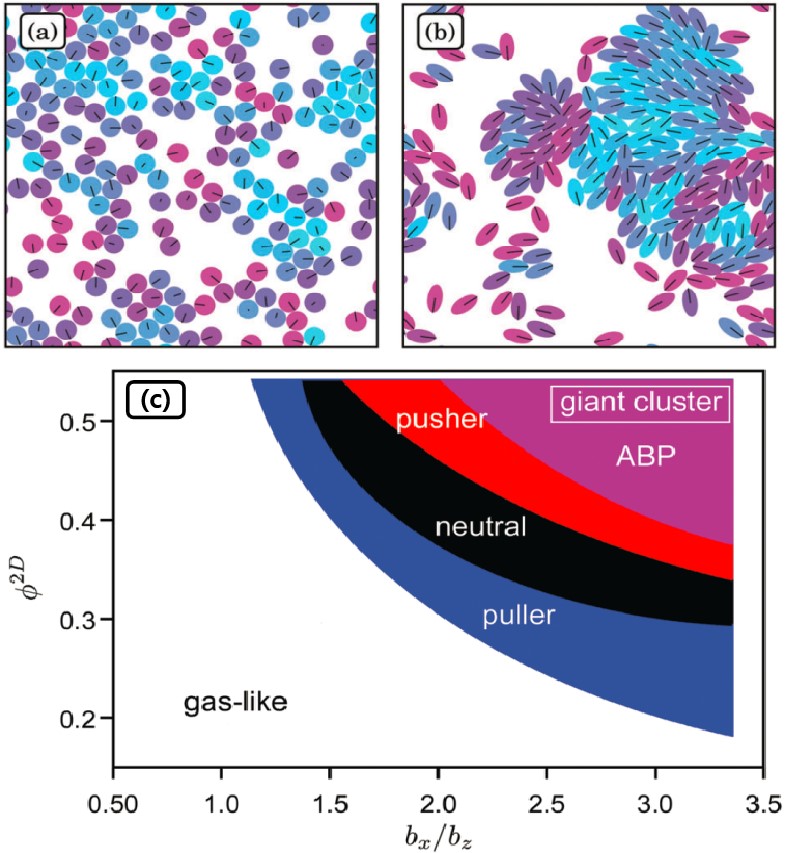}
\caption{Snapshots of the structure of neutral ($\beta=0$) spheroidal squirmers at moderate activity, the area packing fraction $\phi=0.5$, and the aspect ratios (a) $bz/bx=1$ (sphere), (b) 2. Here, the colors indicate the squirmer velocities, as in Fig.~\ref{Fig2}. (c) State diagram for squirmers and ABPs at moderate activity. The blue, black, red, and purple areas indicate giant clusters for pullers ($\beta=1$), neutrals ($\beta=0$), pushers ($\beta=-1$), and active Brownian particles, respectively. Therefore, clusters of pullers appear in all colored areas, clusters of neutral squirmers in the black, red, and purple area, etc.~\cite{theers2018clustering}.
Reproduced with permission.\cite{theers2018clustering} Copyright 2018, Royal Society of Chemistry.
}\label{fig8}
\end{figure}

Related to isotropic spherical swimmers that form clusters due to blocking,
for rod-like swimmers, the alignment from steric interactions is a key mechanism for MIC and MIPS. A recent simulation study
of 3D pusher dumbbell suspension clearly observed activity-induced clustering at modest concentrations~\cite{furukawa2014activity}, where the hydrodynamic trapping of one swimmer by another is thought as the key ingredient for this clustering behavior. A direct comparison between simulations with and without hydrodynamics showed that HIs enhance the dynamic clustering at modest volume fractions, as displayed in Fig.~\ref{fig7}. Furthermore, Theers \emph{et al.}~\cite{theers2018clustering} found a substantial enhancement of MIPS through HI in qusi-2D concentrated systems of spheroidal squirmers with sufficiently large aspect ratios (Fig.~\ref{fig8}), where the squirmers are strongly confined by two boundary walls. This enhancement applies for pushers, pullers, and neutral squirmers, as long as the force dipole is sufficiently weak. A density-aspect-ratio phase diagram (Fig.~\ref{fig8}) for moderate force dipole strength ($\beta$) shows most pronounced phase separation for pullers, followed by neutral squirmers, pushers, and finally ABPs. Here, near-field HIs importantly contribute to the hydrodynamic enhancement
of cluster formation, in contrast to the case of spherical squimers~\cite{matas2014hydrodynamic,yoshinaga2017hydrodynamic}.

\subsection{Self-phoretic swimmers}

\begin{figure}[t]
\includegraphics[width=1.\linewidth]{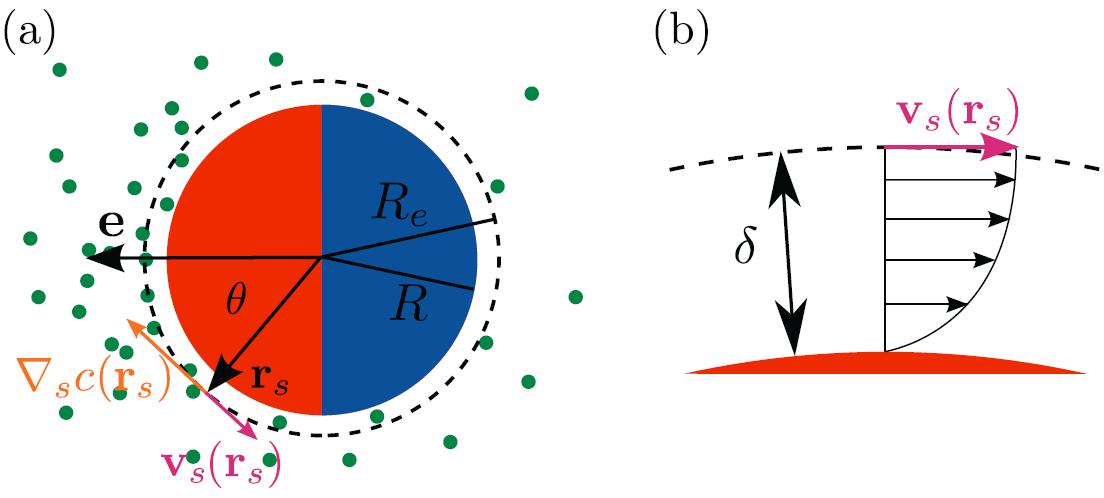}
\caption{(a) Sketch of a self-diffusiophoretic Janus particle of radius $R$ with a catalytic cap (red) and an orientation vector ${\bf e}$ defined toward the cap. The catalytic cap locally creates a tangential concentration gradient of a solute (green dots), $\nabla_s c(r_s)$, at particle surface. The interactions of the particle with the inhomogeneous fluid drive a surface slip flow, ${\bf v}_s(r_s)$. (b) Within the diffusive boundary layer of thickness $\delta$ a surface velocity field develops, and it is proportional to the concentration gradient~\cite{zottl2016emergent}.
Reproduced with permission.\cite{zottl2016emergent} Copyright 2016, IOP Publishing.
}\label{fig9}
\end{figure}

Among diverse artificial microswimmers, one of the important class is the self-phoretic active colloids~\cite{paxton2004catalytic,golestanian2005propulsion,jiang2010active}, which swim due to the phoretic effect driven by self-generated local gradient fields. Phoresis refers to a  {drift} motion of suspended particles in an external gradient field (e.g. electric potential, thermal or chemical gradient, respectively corresponding to electro-, thermo- or diffusiophoresis)~\cite{anderson1989colloid}. Phoretic effect is free of external force and torque, by which active colloids can therefore be straightforwardly realized as long as the particle induces a local gradient by itself. A self-diffusiophoretic Janus particle is employed to explain the basic mechanism underlying the self-phoretic swimmers, as illustrated in Fig.~\ref{fig9}. Briefly, the catalytic cap of a half-coated Janus sphere catalyzes a chemical reaction of the fuel to produce a non-uniform solute concentration around the particle. The inhomogeneous particle-solute interactions drive a surface slip flow around the Janus particle, ${\bf v}_s$, which simultaneously cause the Janus particle to drift parallel to its symmetric axis. By solving the Stokes equation with the slip flow boundary condition, the self-propelled velocity of the Janus particle is determined as ${\bf V}=-\langle{\bf v}_s\rangle$, with the average taken over the particle surface. Depending on the particle-solute interactions, the Janus particle may move along ${\bf e}$ (forward-moving, displaying positive chemotaxis with respect to the reaction product) or against ${\bf e}$ (backward-moving, negative chemotaxis).  {Besides catalytic chemical reaction in solvent, a self-diffusiophoretic Janus particle can also be realized via heating-induced demixing of binary solvent~\cite{bechinger2016active}.}

The self-phoretic swimmers often form dynamic clusters~\cite{bechinger2016active,theurkauff2012dynamic}, where interparticle chemical interactions (chemotaxis) paly an important role~\cite{liebchen2019interactions}. Thus, the squirmers or dumbbells with phantom flagella, as introduced above, are not suitable to model self-phoretic swimmers, since they do not describe the gradient fields. In order to properly mimic the self-phoretic colloids, particle-based (explicit) mesoscale simulation methods have been employed to study the self-diffusiophoretic and self-thermophoretic swimmers~\cite{ruckner2007chemically,yang2011simulations,yang2014hydrodynamic,huang2016microscopic}, in which thermal fluctuations, hydrodynamics, heat conduction and mass diffusion can be correctly captured.

\begin{figure}
\includegraphics[width=1.\linewidth]{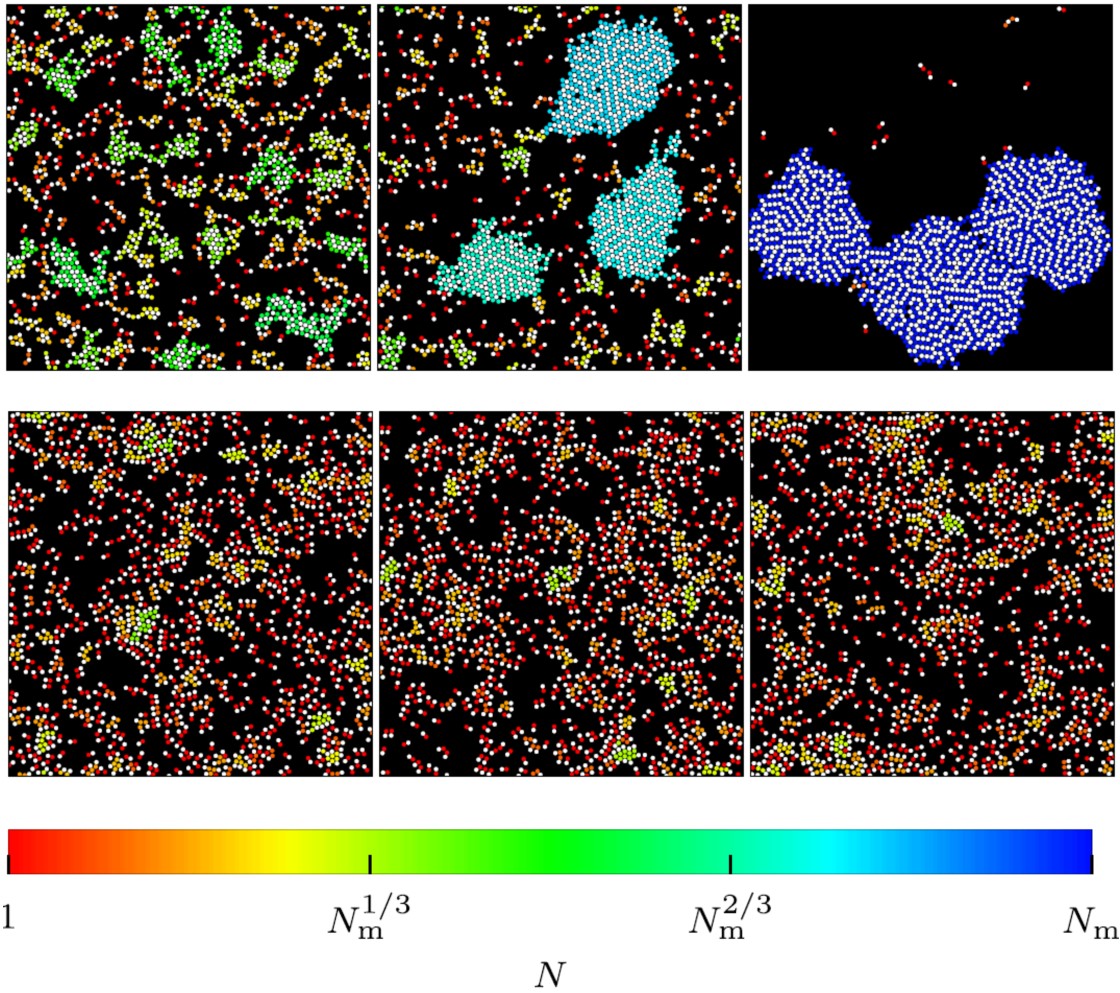}
\caption{Instantaneous configurations of $N_m=1000$ forward-moving (top)
and backward-moving self-diffusiophoretic dimers (bottom) for area fraction
$\phi=0.2$ at short, intermediate, and long times (from left to right). Here, the two constituent beads of the dimer have the same size. Clusters of N swimmers are colored to a logarithmic scale~\cite{colberg2017many}.
Reproduced with permission.\cite{colberg2017many} Copyright 2017, AIP Publishing.
}\label{fig10}
\end{figure}

Recently, Kapral and coworkers~\cite{colberg2017many} performed large-scale explicit mesoscale simulation to study the phase behavior of self-diffusiophoretic dimers, whose one bead catalyzes chemical reaction. The dimer dynamics is confined to a 2D plane, although the fluid flow and concentration fields are three dimensional. They found that the chemotactic effects due to chemical gradients are the dominant factor in the collective behavior of the active system, at the same time, which is also influenced by the dimer geometry and HI. Forward-moving dimers (positive chemotaxis) usually segregate into high and low density phases, while backward-moving dimers (negative chemotaxis) keep globally homogeneous states with strong fluctuations, as shown in Fig.~\ref{fig10}. Further, they investigated the dynamic cluster states in a 3D suspension of self-diffusiophoretic Janus spheres~\cite{huang2017chemotactic}. Depending on microscopic characteristics, chemotactic and hydrodynamic interactions can operate either cooperatively or against one another to enhance or suppress dynamical clustering. The particle aggregation depend strongly on the size of the catalytic cap on the active Janus spheres. Particularly, when eliminating chemotactic interactions, the cluster of Janus motors gradually breaks apart.

\begin{figure}[h]
\includegraphics[width=1.\linewidth]{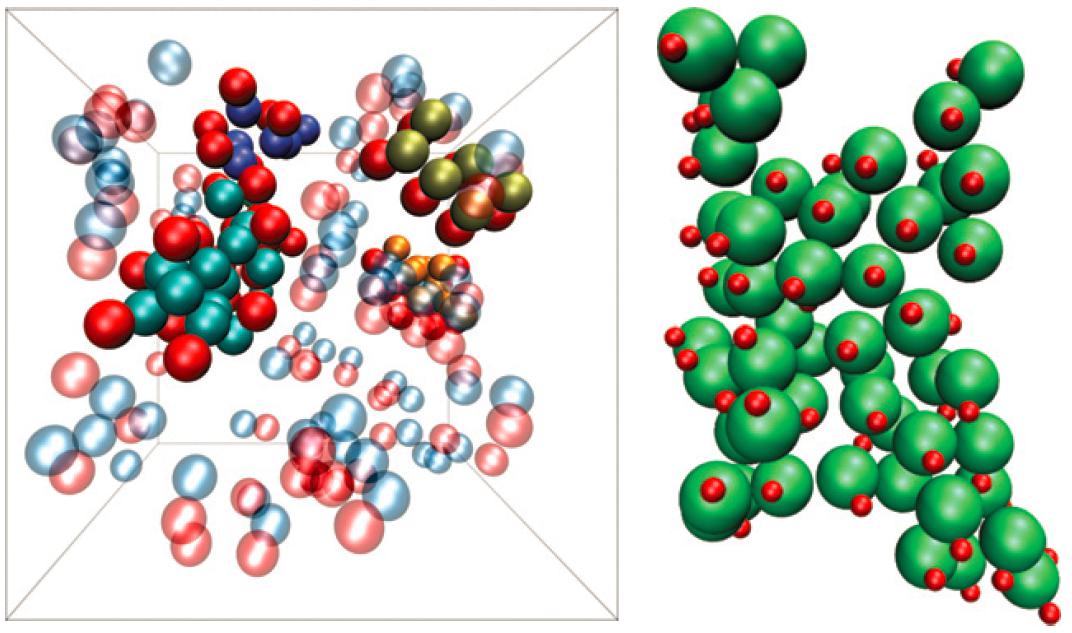}
\caption{Snapshots of ensembles of self-thermophoretic
dimers, with red heated beads. (Left) 100 symmetric swimmers (the two beads of the dimer having equal size) with volume fraction $\phi=0.05$. Nonassembled dimers are translucent, and dimers assembled in clusters are solid. (Right) close-up snapshots of large cluster in a simulation with 500 asymmetric swimmers~\cite{wagner2017hydrodynamic}.
Reproduced with permission.\cite{wagner2017hydrodynamic} Copyright 2017, IOP Publishing.
}\label{fig11}
\end{figure}

Using a similar mesoscale scheme, Wagner \emph{et al.}~\cite{wagner2017hydrodynamic} simulated a system of self-thermophoretic dimeric swimmers, where one bead of the dimer can heat the surrounding solvent and create a local thermal gradient. Although the dimers are thermophobic (negative thermotaxis), they can still form a cluster due to HIs. In particular, the interplay of attractive hydrodynamics with negative thermotaxis leads to the formation of swarming clusters with a flattened geometry in the three-dimensional system (Fig.~\ref{fig11}). These results suggest that for the present self-phoretic colloids the clustering mechanism primarily arises from phoretic and hydrodynamic interactions, instead of the motility-induced clustering.

\section{Concluding Remarks}\label{conclude}
In this article, we reviewed the recent progress on the study of collective assembly of active colloids by using analytical theory and computer simulations. We first briefly reviewed the studies on the MIPS of repulsive spherical ABP systems, which has been one of major focuses in the past decade in the active matter community, followed by the effect of shape anisotropy of active colloids and the hydrodynamic interactions on emergent behaviour of active colloids.  So far, for the simplified overdamped repulsive active Brownian particles, the MIPS has been reasonably well understood in the framework of an equilibrium-like mean field theory. However, increasing amounts of new evidence have shown that the equilibrium-like mean field picture may break down with some small change on the dynamics. For example, it was found that in underdamped active Brownian particle systems, the effect of inertia can induce a re-entrant melting transition with increasing activity, which can not be explained in any present mean field theory for MIPS~\cite{MIPS_lowen_2019}. Moreover, in experiments, no ABP can be a perfect sphere, and the shape anisotropy can induce a torque on ABPs. It was recently found that in system of circle active hard spheres, i.e., ABP with a torque~\cite{mazhan2017}, the dynamic ground state of the fluid phase is hyperuniform, which is fundamentally different from conventional equilibrium fluids with short range interactions~\cite{leihu2019sa,leihu2019pnas}. These suggest that we are far from fully understanding MIPS, and there are still many open questions to be addressed.

\begin{acknowledgements}
This work is supported by Nanyang Technological University Start-Up Grant M4081781.120; Academic Research Fund from Singapore Ministry of Education Grant MOE2019-T2-2-010 and RG104/17 (S); and Advanced Manufacturing and Engineering Young Individual Research Grant (A1784C0018) by the Science and Engineering Research Council of Agency for Science, Technology and Research Singapore. M.Y. acknowledges support from National Natural Science Foundation of China (No. 11874397, 11674365).
\end{acknowledgements}

\bibliographystyle{h-physrev}
%\bibliography{reference}
%printbibliography

\end{document}